%% file: draft_ac_final_3_arxiv.tex
\documentclass[conference,a4paper]{IEEEtran}
\usepackage{graphicx} 
\usepackage{array}
\usepackage{arydshln}
\usepackage{xcolor,colortbl}
\usepackage{mleftright}
\usepackage{cite}
\usepackage{enumerate}
\usepackage{tikz}

\usepackage{mathtools}
\input{math-macros}

\usepackage{setspace}
\renewcommand{\opP}{\operatorname{P}}

\newcommand{\diam}{\operatorname{diam}}

\allowdisplaybreaks[3]
\mathtoolsset{showonlyrefs=true}   
\usetikzlibrary{matrix,patterns,calc,decorations.fractals,arrows,external}
 
\definecolor{tblblue}{rgb}{0.93,0.93,1.0}
\definecolor{tblred}{rgb}{1,0.93,0.93}
\definecolor{darkblue}{rgb}{0,0,0.7} 
\definecolor{darkgreen}{RGB}{20,120,43} 
\definecolor{darkred}{rgb}{0.8,0,0} 
\definecolor{lightblue}{RGB}{101,124,191}
\definecolor{skyblue}{RGB}{135,206,235}
\definecolor{gold}{RGB}{204,168,66}
\definecolor{strongblue}{RGB}{60,146,228}
\definecolor{lightgray}{gray}{0.5}
\definecolor{verylightgray}{RGB}{101,124,191}
\definecolor{mistyrose}{RGB}{238,213,210}
\definecolor{firebrick3}{RGB}{205,38,38} 
 
\author{\IEEEauthorblockN{
Giovanni Alberti$^\ddagger$,
Helmut B\"olcskei$^\dagger$,
Camillo De Lellis$^\ast$, 
G\"unther Koliander$^\diamond$,   
and 
Erwin Riegler$^\dagger$
\medskip}
\IEEEauthorblockA{$^\ddagger$Dept.~of~Mathematics, University of Pisa, Italy, 
Email: galberti1@dm.unipi.it}
\IEEEauthorblockA{$^\dagger$Dept.~of~IT~\&~EE, ETH Zurich, Switzerland, 
Email: \{boelcskei, eriegler\}@nari.ee.ethz.ch}
\IEEEauthorblockA{$^\ast$Dept.~of~Mathematics, University of Zurich, Switzerland, 
Email: camillo.delellis@math.uzh.ch}
\IEEEauthorblockA{$^\diamond$Inst.~of~Telecommunications, TU Wien, Austria,
Email: 	guenther.koliander@tuwien.ac.at}
}

\begin{document}

\title{Lossless Linear Analog Compression}

\maketitle

\begin{abstract}
We establish the fundamental limits of lossless linear analog compression by considering the recovery of random vectors $\rvecx\in\reals^m$  from the noiseless linear measurements $\rvecy=\matA\rvecx$ with measurement matrix $\matA\in\reals^{n\times m}$. 
Specifically, for a random vector $\rvecx\in\reals^{m}$ of \emph{arbitrary} distribution we show that $\rvecx$ can be recovered with zero error probability from 
$n>\inf\underline{\dim}_\mathrm{MB}(\setU)$ linear measurements, where $\underline{\dim}_\mathrm{MB}(\cdot)$ denotes the lower modified Minkowski dimension and the infimum is over all sets $\setU\subseteq \reals^{m}$ with $\opP[\rvecx\in\setU]=1$. This achievability statement holds for Lebesgue almost all measurement matrices $\matA$. 
We then show that \emph{$s$-rectifiable} random vectors---a stochastic generalization of $s$-sparse vectors---can be recovered
with zero error probability from $n> s$ linear measurements.
From classical compressed sensing theory 
we would expect $n\geq s$ to be necessary for successful recovery of $\rvecx$.  
Surprisingly, certain classes of $s$-rectifiable random vectors
can be recovered from fewer than $s$ measurements. 
Imposing an additional regularity condition on the distribution of $s$-rectifiable random vectors $\rvecx$, 
we do get the expected converse result of $s$ measurements being necessary. 
The resulting class of random vectors appears to be new and will be referred to as \emph{$s$-analytic} random vectors.   
\end{abstract}

\section{Introduction}

Compressed sensing \cite{do06,carota06,fora13} deals with the recovery of unknown sparse vectors $\vecx\in\reals^m$ from a small (relative to $m$) number, $n$, of linear measurements of the form $\vecy=\matA\vecx$, where $\matA\in\reals^{n\times m}$ is referred to as the measurement matrix. 
Wu and Verd\'u \cite{wuve10,wuve12} developed an information-theoretic framework for compressed sensing, fashioned as an almost lossless analog compression problem. 
Specifically, \cite{wuve10} presents asymptotic achievability bounds, which show that for almost all (a.a.) measurement matrices $\matA$ a random i.i.d. vector $\rvecx$ can be recovered with arbitrarily small probability of error from $n=\lfloor R m\rfloor$ 
linear measurements, provided that $R>R_\mathrm{B}$, where $R_\mathrm{B}$ denotes the Minkowski dimension compression rate \cite[Def. 10]{wuve10} of $\rvecx$. 
For the special case of the i.i.d. components in $\rvecx$ having a discrete-continuous mixture distribution, this threshold is tight in the sense of $R\geq R_\mathrm{B}$ being necessary for the existence of a measurement matrix $\matA$ such that $\rvecx$ can be recovered with probability of error strictly smaller than $1$ for $m$ sufficiently large. 
Discrete-continuous mixture distributions $\rho\mu^\mathrm{c}+(1-\rho)\mu^\mathrm{d}$ are relevant as $\lfloor\rho m\rfloor$---by the law of large numbers---can be interpreted as the sparsity level of $\rvecx$ and $R_\mathrm{B}=\rho$.  A more direct and non-asymptotic (i.e., fixed-$m$) statement in  \cite{wuve10} says that a.a. (with respect to a $\sigma$-finite Borel measure) $s$-sparse random vectors can be recovered with zero probability of error provided that $n>s$. 
Again, this result holds for Lebesgue a.a. measurement matrices $\matA\in\reals^{n\times m}$. A corresponding converse does, however, not seem to be available.
 
\emph{Contributions.} We establish the fundamental limits of lossless (i.e., zero probability of error) linear analog compression in the non-asymptotic (i.e., fixed-$m$) regime
for random vectors $\rvecx$ of arbitrary distribution. In particular, $\rvecx$ need not be i.i.d. or  
supported on  the union of subspaces (as in classical compressed sensing theory). 
The formal statement of the problem we consider is as follows. 
Suppose we have $n$ (noiseless) linear measurements of the random vector $\rvecx\in\reals^m$ in the form of $\rvecy=\matA\rvecx$.   
For a given $\varepsilon\in[0,1)$, we want to determine whether a decoder, i.e., a Borel measurable map $g_\matA\colon\reals^n\to\reals^m$ exists such that  
\begin{align}\label{eq:decoder}
\opP\mleft[g_\matA\big(\matA\rvecx\big)\neq\rvecx\mright]\leq\varepsilon. 
\end{align}
Specifically, we shall be interested in statements of the following form:\\  
\vspace*{-4truemm}

\noindent\emph{Property P1:} For Lebesgue a.a. measurement matrices $\matA\in\reals^{n\times m}$, there exists a decoder $g_\matA$ satisfying \eqref{eq:decoder} with $\varepsilon=0$.\vspace*{1mm}\\
\noindent\emph{Property P2:}
There exist an $\varepsilon\in[0,1)$, an  $\matA\in \reals^{n\times m}$, and a decoder $g_\matA$ satisfying  \eqref{eq:decoder}.\\
\vspace*{1mm}

\vspace*{-4truemm}
\noindent
Our main achievability result is as follows. 
For $\rvecx\in\reals^{m}$ of arbitrary distribution, we show that P1 holds for   
$n>\inf\underline{\dim}_\mathrm{MB}(\setU)$,  where $\underline{\dim}_\mathrm{MB}(\cdot)$ denotes the lower modified Minkowski dimension (see Definition \ref{dfndimlocal}) and the infimum is over all sets $\setU\subseteq \reals^{m}$ with $\opP[\rvecx\in\setU]=1$. 
We emphasize that it is the usage of modified Minkowski dimension, as opposed to Minkowski dimension, that allows us to obtain an achievability result for $\varepsilon=0$. 
The central conceptual element in the proof of this statement is a slightly modified version of the probabilistic null-space property first reported in \cite{striagbo15}.
The asymptotic achievability bounds in  \cite{wuve10} can be recovered in our framework.

We make the connection of our results to classical compressed sensing explicit by considering random vectors $\rvecx\in\reals^m$ that
consist of $s$ i.i.d. Gaussian entries at positions drawn uniformly at random and that have all other entries equal to zero.
This class can be considered a stochastic analogon of $s$-sparse vectors and belongs to the more general class of $s$-rectifiable random vectors, originally introduced in \cite{kopirihl15} to derive  a new concept of entropy that goes beyond classical entropy and differential entropy.
Specifically, a random vector $\rvecx$ is said to be $s$-rectifiable if there exists an $s$-rectifiable set $\setU$ \cite[Def. 4.1]{de00} with $\opP[\rvecx\in\setU]=1$ and the distribution of $\rvecx$ is absolutely continuous with respect to the $s$-dimensional Hausdorff measure.\footnote{Note that the classical Lebesgue decomposition  of measures into continuous, discrete, and singular parts is not useful for $s$-rectifiable random vectors as their distributions are always singular (except for the trivial cases $s=m$ and $s=0$).
We therefore use the $s$-dimensional Hausdorff measure as reference measure for the ambient space.} 
Our achievability result particularized for $s$-rectifiable random vectors shows that P1 holds for $n>s$. 
From classical compressed sensing theory 
we would expect $n\geq s$ to be necessary for successful recovery of $\rvecx$.
Our information-theoretic framework reveals, however, that this is not the case for certain classes of $s$-rectifiable random vectors. This will be illustrated by way of an example, which constructs a $2$-rectifiable set $\setG\subseteq \reals^3$ of positive $2$-dimensional Hausdorff measure  that can be compressed linearly in a one-to-one fashion into $\reals$. 
Operationally, this implies that zero error probability recovery from $n=1<s=2$ measurement is possible. 
What renders this result  surprising is that $\setG$ contains the image---under a continuous differentiable mapping---of a set in $\reals^2$ of positive Lebesgue measure. 
We then show that imposing a regularity condition on the distribution of $\rvecx$, does lead to the expected converse result in the sense of $n\geq s$ being necessary for P2 to hold. 
The resulting class of random vectors appears to be new and will be referred to as $s$-analytic random vectors. 

\emph{Notation.}
Capital boldface letters $\matA,\matB,\dots$  designate deterministic matrices and lower-case boldface letters $\veca,\vecb,\dots$ stand for deterministic vectors. 
We use sans-serif letters, e.g.\ $\rvecx$, for random quantities and roman letters, e.g. $\vecx$, for deterministic quantities. 
For measures $\mu$ and $\nu$ on the same measurable space, we write $\mu\ll\nu$ to express that $\mu$ is absolutely continuous with respect to $\nu$ (i.e., for every measurable set $\setA$, $\nu(\setA)=0$ implies $\mu(\setA)=0$). 
The product measure of  $\mu$ and $\nu$ is denoted by $\mu\times\nu$. 
The superscript  $\tp{}$ stands for transposition. 
$\|\vecx\|_2=\sqrt{\tp{\vecx}\vecx}$ is the Euclidean norm of $\vecx$ and $\|\vecx\|_0$ denotes the number of non-zero entries of $\vecx$.  
For the Euclidean space $(\reals^k,\|\cdot\|_2)$, we let the open ball of radius $\rho$ centered at $\vecu\in \reals^k$ be $\setB_k(\vecu,\rho)$, and  $V(k,\rho)$  refers to its volume. 
$\mathscr{L}^{n}$ denotes the Lebesgue measure on $\reals^n$. 
If $f\colon\reals^k\to\reals^l$ is differentiable, we write $Df(\vecv)$ for the differential of $f$ at $\vecv\in\reals^k$ and we define the $\min(k,l)$-dimensional Jacobian $Jf(\vecv)$ at $\vecv\in\reals^k$ by 
$Jf(\vecv)=\sqrt{\det(Df(\vecv)\tp{(Df(\vecv))})}$, if $l<k$, and $Jf(\vecv)=\sqrt{\det(\tp{(Df(\vecv))}Df(\vecv))}$, if $l\geq k$. 
For a mapping $f$, 
$f\not\equiv \veczero$ means that $f$ is not identically zero. 
For $f\colon\reals^k\to\reals^l$ and $\setA\subseteq\reals^k$, $f|_\setA$ denotes the restriction of $f$ to $\setA$. A mapping is said to be $C^1$ if its  derivative exists and is continuous. $\ker(f)$ stands for the kernel of $f$.  

The definitions of the fractal quantities used in this paper are standard and can be found,  along with their basic properties in, e.g.,  \cite{fa90,amfupa00}. Throughout the paper, we omit proofs due to space limitations. 
 
\section{Achievability}

We quantify the description complexity of random vectors $\rvecx\in\reals^m$ of general distribution through the infimum over the lower modified Minkowski dimensions of sets $\setU\subseteq\reals^{m}$ with  $\opP[\rvecx\in\setU]=1$. 
We start by defining Minkowski dimension.
 
\begin{dfn}(Minkowski dimension\footnote{Minkowski dimension is sometimes also referred to as box-counting dimension, which is the origin of the subscript B in the notation $\dim_\mathrm{B}(\cdot)$ used below.})\label{dfndim}
Let $\setU$ be a non-empty bounded set in $\reals^{m}$.  
The lower Minkowski dimension of $\setU$ is defined as 
\begin{align}
\underline{\dim}_\mathrm{B}(\setU)=\liminf_{\rho\to 0} \frac{\log N_\setU(\rho)}{\log \frac{1}{\rho}}\nonumber
\end{align}
and the upper Minkowski dimension as  
\begin{align}
\overline{\dim}_\mathrm{B}(\setU)=\limsup_{\rho\to 0} \frac{\log N_\setU(\rho)}{\log \frac{1}{\rho}},\nonumber 
\end{align}
where 
\begin{align}
N_\setU(\rho)=\min\Big\{k \in\naturals : \setU\subseteq\hspace*{-4truemm} \bigcup_{i\in\{1,\dots,k\}}\hspace*{-4truemm} \setB_{m}(\vecu_i,\rho),\ \vecu_i\in \reals^{m}\Big\}\nonumber
\end{align}
is the covering number of $\setU$ for radius $\rho$.
If $\underline{\dim}_\mathrm{B}(\setU)=\overline{\dim}_\mathrm{B}(\setU)=: \dim_\mathrm{B}(\setU)$, we  say that $\dim_\mathrm{B}(\setU)$ is \emph{the} Minkowski dimension of $\setU$.
\end{dfn}

Minkowski dimension is a useful measure only for (non-empty) bounded sets, as it equals infinity for unbounded sets.
A measure of description complexity that applies to unbounded  sets as well is modified Minkowski dimension. 

\begin{dfn}(Modified Minkowski dimension)\label{dfndimlocal} Let $\setU\subseteq\reals^{m}$ be a non-empty  set. The lower modified Minkowski dimension of $\setU$ is defined as  
\begin{align}
\underline{\dim}_\mathrm{MB}(\setU)=
\inf\mleft\{\sup_{i\in\setI} \underline{\dim}_\mathrm{B}(\setU_i) : \setU\subseteq \bigcup_{i\in\setI}\setU_i\mright\},
\end{align}
where the infimum is over all countable covers of $\setU$ by non-empty bounded Borel sets. 
The upper modified Minkowski dimension of $\setU$ is 
\begin{align}
\overline{\dim}_\mathrm{MB}(\setU)=
\inf\mleft\{\sup_{i\in\setI} \overline{\dim}_\mathrm{B}(\setU_i) : \setU\subseteq \bigcup_{i\in\setI}\setU_i\mright\},
\end{align}
where, again, the infimum is over all countable covers of $\setU$ by non-empty bounded Borel sets. 
If $\underline{\dim}_\mathrm{MB}(\setU)=\overline{\dim}_\mathrm{MB}(\setU)=: \dim_\mathrm{MB}(\setU)$, we  say that $\dim_\mathrm{MB}(\setU)$ is \emph{the} modified Minkowski dimension of $\setU$.
\end{dfn}
Upper modified Minkowski dimension has  the advantage of being countably stable \cite[Sec. 3.4]{fa90}, whereas upper Minkowski dimension is only finitely stable. 
For example, all countable subsets of $\reals^{m}$ have modified Minkowski dimension zero, but there are countable subsets of $\reals^{m}$ with nonzero Minkowski dimension: 
\begin{exa}\cite[Ex. 3.5]{fa90}\label{exa:MBB}
Let $\setF=\{0,1/2,1/3,\dots\}$. Then, ${\dim}_\mathrm{MB}(\setF)=0<{\dim}_\mathrm{B}(\setF)=1/2$.
\end{exa}  
The fact that upper modified Minkowski dimension is countably stable will turn out to be of key importance in particularizing our achievability result, stated next, for $s$-rectifiable random vectors.   
\begin{thm}\label{th1}
For $\rvecx\in\reals^{m}$ of arbitrary distribution,   
$n>\inf\underline{\dim}_\mathrm{MB}(\setU)$ is sufficient for Property P1 to hold, where the infimum is over all 
sets $\setU\subseteq \reals^{m}$ with $\opP[\rvecx\in\setU]=1$. 
\end{thm}

This theorem generalizes the  achievability result of \cite{wuve10} to random vectors $\rvecx\in\reals^m$ of \emph{arbitrary} distribution. 
Specifically, neither do the entries of $\rvecx$ have to be i.i.d. nor does $\rvecx$ have to be generated according to the finite union of subspaces model. 
Finally, perhaps most importantly, the result is non-asymptotic (i.e., for finite $m$) and pertains to zero error probability. 

The central conceptual element in the derivation of Theorem~\ref{th1} is the following probabilistic null-space property, first reported in \cite{striagbo15} for (non-empty) bounded sets and expressed in terms of lower Minkowski dimension. 
If the lower modified Minkowski dimension of a non-empty (possibly unbounded) set $\setU$ is smaller than $n$, then,  for a.a.  measurement matrices $\matA$, the set $\setU$ intersects the $(m-n)$-dimensional kernel of $\matA$  at most trivially. 
What is remarkable here is that the notions of Euclidean dimension (for the kernel of the mapping) and of lower modified  Minkowski dimension (for $\setU$) are compatible.  
The formal statement is as follows.  

\begin{prp}\label{prp:ns}
Suppose that $\setU\subseteq\reals^m$ with $\underline{\dim}_\mathrm{MB}(\setU)<n$. 
Then, we have 
\begin{align}
\ker(\matA)\cap(\setU\!\setminus\!\{\matzero\})=\emptyset 
\end{align}
for Lebesgue a.a. matrices $\matA\in\reals^{n\times m}$. 
\end{prp}


We next particularize our achievability result for $s$-rectifiable random vectors $\rvecx$---defined below---and
start by introducing the central concepts needed, namely, Hausdorff measures, Hausdorff dimension, and (locally) Lipschitz mappings. 
 $s$-rectifiable random vectors are important as they constitute a stochastic analogon of the union of subspaces
model used pervasively in classical compressed sensing theory.

\begin{dfn}(Hausdorff measure)\label{dfnH}
Let  $s\in [0,\infty)$ and $\setU\subseteq \reals^{m}$. 
The $s$-dimensional Hausdorff measure of $\setU$ is given by 
\begin{align}
\mathscr{H}^{s}(\setU)=\lim_{\delta\to 0}\mathscr{H}_\delta^{s}(\setU)
\end{align}  
where, for $0<\delta\leq \infty$, 
\begin{align}
&\mathscr{H}_\delta^{s}(\setU)\\
&=\frac{V(s,1)}{2^s}\inf\mleft\{\sum_{i\in\setI}\diam(\setU_i)^s : \diam(\setU_i)<\delta, \setU\subseteq\bigcup_{i\in\setI}\setU_i\mright\}
\end{align}
for countable covers $\{\setU_i\}_{i\in\setI}$ and the diameter of $\setU\subseteq\reals^{n}$ is defined as
\begin{align}
\diam(\setU)=
\begin{cases}
\sup\{\|\vecu-\vecv\|_2 : \vecu,\vecv\in\setU\},&\text{for}\ \setU\neq\emptyset\\
0,&\text{for}\ \setU=\emptyset.
\end{cases}
\end{align}
\end{dfn} 
\begin{dfn}(Hausdorff dimension)
The Hausdorff dimension of $\setU\subseteq\reals^{m}$ is
\begin{align}
\dim_{\mathrm{H}}(\setU)
&=\sup\{s\geq 0 : \mathscr{H}^{s}(\setU)=\infty\}\\
&=\inf\{s\geq 0 : \mathscr{H}^{s}(\setU)=0\}\label{eq:defhd}, 
\end{align} 
i.e., $\dim_{\mathrm{H}}(\setU)$  is the value of $s$ for which the sharp transition  from $\infty$ to $0$ occurs in Figure \ref{fig:hd}.
\end{dfn}

\begin{figure}
\begin{center}
\begin{tikzpicture}[scale=1.5]
    \draw [thick, <->] (0,2) node (yaxis) [above] {$\mathscr{H}^{s}(\setU)$}
        |- (4,0) node (xaxis) [right] {$s$};
    \draw [ultra thick] (0,1.6) coordinate (a_1) -- (1.8,1.6) coordinate (a_2);
    \draw [ultra thick] (1.8,0) coordinate (b_1) -- (3,0) coordinate (b_2);
    \node[left] at (a_1) {$\infty$};
    \node[left] at (0,0) {$0$};
    \node[below] at (b_1) {$\dim_{\mathrm{H}}(\setU)$};
    \node[below] at (b_2) {$m$};
\end{tikzpicture}
\caption{(\!\!\cite[Fig. 2.3]{fa90}) 
Graph of  $\mathscr{H}^{s}(\setU)$ as a function of $s\in[0,m]$ for a set $\setU\subseteq\reals^{m}$. 
\label{fig:hd}} 
\end{center}
\vspace*{-3truemm}
\end{figure}

\begin{dfn}(Locally Lipschitz mapping)
\begin{enumerate}[(i)]
\item
A mapping $f\colon\setU\to \reals^{l}$, where $\setU\subseteq\reals^{k}$, is Lipschitz if there exists a constant $L \geq 0$ such that 
\begin{align}\label{eq:lip}
\|f(\vecu)-f(\vecv)\|_2\leq L\|\vecu-\vecv\|_2,
\end{align}
for all $\vecu,\vecv\in\setU$. 
The smallest constant $L$ for which \eqref{eq:lip} holds is called the Lipschitz constant of $f$;
\item
a mapping $f\colon\reals^{k}\to \reals^{l}$ is locally Lipschitz if, for each compact set $\setK\subseteq \reals^{k}$, the mapping $f|_\setK\colon\setK\to\reals^l$ is Lipschitz.  
\end{enumerate}
\end{dfn}

We are now ready to define the notion of $s$-rectifiable sets and $s$-rectifiable random vectors.
 
\begin{dfn}\label{dfn:recset}
An $\mathscr{H}^{s}$-measurable set $\setU\subseteq\reals^{m}$  is called $s$-rectifiable if there exist a countable  set $\setI$, bounded Borel sets $\setA_i\subseteq\reals^s$, $i\in\setI$, and  Lipschitz  mappings $\varphi_i\colon\setA_i\to \reals^{m}$, $i\in\setI$ such that 
\begin{align}
\mathscr{H}^{s}\Big(\setU\setminus\bigcup_{i\in\setI}\varphi_i(\setA_i)\Big)=0. 
\end{align}
\end{dfn}


\begin{dfn}\label{dfnXrec}
The random vector $\rvecx\in\reals^{m}$ is called  $s$-rectifiable if there exists an $s$-rectifiable 
set $\setU\subseteq \reals^{m}$ with $\opP[\rvecx\in\setU]=1$ and  $\mu_\rvecx\ll \mathscr{H}^{s}$. 
\end{dfn}

The following example speaks to the relevance of the notion of $s$-rectifiable random vectors.

\begin{exa}\label{exrecana1a}
Suppose that $\rvecx\in\reals^{m}$ has  $s$ i.i.d. Gaussian entries  at positions drawn uniformly at random and all other entries are equal to zero. 
Then,  the $s$-rectifiable  set 
\begin{align}\label{eq:setS}
\setU&=\{\vecx\in\reals^m : \|\vecx\|_0=s\},
\end{align}
satisfies   $\opP[\rvecx\in\setU]=1$. 
We show in Example \ref{exrecana1} that $\mu_\rvecx\ll \mathscr{H}^{s}$,
which implies $s$-rectifiability of $\rvecx$.
\end{exa}

We next establish an important uniqueness property of $s$-rectifiable random vectors.

\begin{lem}\label{lem:uniques}
If $\rvecx\in\reals^m$  is $s$-rectifiable and $t$-rectifiable, then  $s=t$. 
\end{lem}

Roughly speaking the reason for this uniqueness is the following. 
If we reduce $s$, then there exists no $s$-rectifiable set $\setU$ with $\opP[\rvecx\in\setU]=1$, 
if we increase it, then $\mu_\rvecx\ll \mathscr{H}^{s}$ is violated as a consequence of the sharp transition behavior of Hausdorff measure depicted in Figure \ref{fig:hd}. 

We next particularize our achievability result, Theorem \ref{th1}, for $s$-rectifiable random vectors. 
To this end, we first establish an auxiliary result. 

\begin{lem}\label{lem:MBrec}
Each $s$-rectifiable random vector $\rvecx\in\reals^m$ has at least one set $\setU\subseteq\reals^m$ with  $\opP[\rvecx\in\setU]=1$ and $\overline{\dim}_\mathrm{MB}(\setU)\leq s$.
\end{lem}
Combining Lemma \ref{lem:MBrec} and Theorem \ref{th1} yields the following achievability result for $s$-rectifiable random vectors. 
\begin{cor}\label{cor:rectifiable}
For $\rvecx\in\reals^{m}$ $s$-rectifiable, $n>s$ is sufficient for P1 to hold. 
\end{cor}

\section{Converse}\label{sec:analytic}

\begin{figure}\centering
\includegraphics[width=0.45\textwidth]{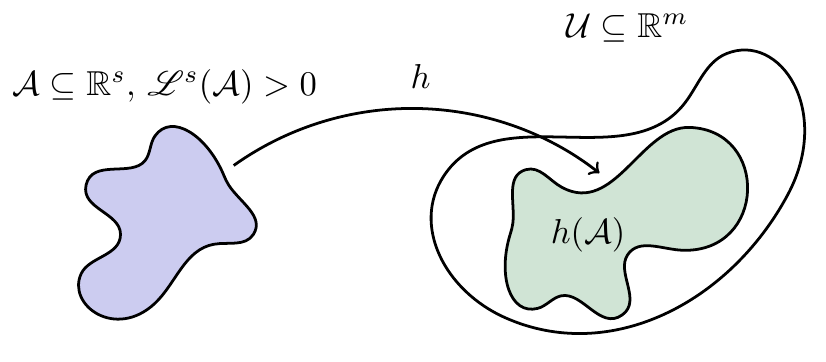}
\caption{\small The set $\setU\subseteq\reals^m$  contains the image of a set $\setA\subseteq\reals^s$ with positive Lebesgue measure $\mathscr{L}^{s}(\setA)>0$. The mapping $h$ is one-to-one on $\setA$.\label{fig:2}}
\end{figure}

Our achievability result particularized for $s$-rectifiable random vectors shows that P1 holds for $n>s$. 
From classical compressed sensing theory we would expect $n\geq s$ to be necessary for successful recovery of $\rvecx$. 
Our information-theoretic framework reveals, however, that this is not the case for \emph{certain} classes of $s$-rectifiable random vectors. This surprising phenomenon will be illustrated through the following example.
We construct a  \emph{$2$-rectifiable} set $\setG\subseteq \reals^3$ of positive $2$-dimensional Hausdorff measure  that can be compressed linearly in a one-to-one fashion into $\reals$. 
What renders this result surprising is that all this is possible although $\setG$ contains the one-to-one image---under a continuous differentiable mapping---of a set in $\reals^2$ of positive Lebesgue measure (see Figure \ref{fig:2}). Operationally, this shows that $2$-rectifiable random vectors $\rvecx$ with $\opP[\rvecx\in\setG]=1$  can be recovered from $n=1 < s=2$ linear measurement with zero probability of error. Let us proceed to the formal statement of the example. 

\begin{exa}\label{exa.counter}
We construct a $2$-rectifiable set $\setG\subseteq\reals^3$ with $\mathscr{H}^{2}(\setG)>0$  and a corresponding linear mapping $f\colon\reals^3\to\reals$ such that $f$ is one-to-one on $\setG=h(\setA)$, where $h\colon\reals^2\to\reals^3$ is $C^1$, $\setA\subseteq\reals^2$ has $\mathscr{L}^{2}(\setA)>0$, and $h$ is one-to-one on $\setA$. 

\emph{Construction of $\setG$}: 
It can be shown that there exist a $C^1$-mapping $\kappa\colon\reals^2\to\reals$ and a bounded Borel set $\setA\subseteq\reals^2$ with $0<\mathscr{L}^{2}(\setA)<\infty$ such that $\kappa$ is one-to-one on $\setA$. Let $\setG=\{\tp{(\vecz\ \kappa(\vecz))}\mid \vecz\in\setA\}\subseteq\reals^3$. 
Since $\kappa$ is a  $C^1$-mapping, 
\begin{align}
h\colon\reals^2&\to\reals^3\\
\vecz& \mapsto\tp{(\vecz\ \kappa(\vecz))}
\end{align}
is locally Lipschitz. 
We then cover $\reals^2$ by compact sets $\setK_i$, $i\in\setI$, with $\setI$ countable. 
The local Lipschitz property of $h$ implies that the mappings $\varphi_i=h\!\!\mid_{\setK_i}$, $i\in\setI$, are Lipschitz. 
Therefore,  by Definition \ref{dfn:recset}, 
\begin{align}
\setG=\bigcup_{i\in\setI}\varphi_i(\setK_i\cap\setA)
\end{align}
is $2$-rectifiable. 

\emph{$\mathscr{H}^{2}(\setG)>0$}:
Let $\pi\colon\reals^3\to\reals^2$, $\tp{(x_1\ x_2\ x_3)}\to \tp{(x_1\ x_2)}$. Clearly, $\pi$ is a Lipschitz mapping with Lipschitz constant equal to one. 
Using \cite[Prop. 2.49, Property (iv)]{amfupa00} and \cite[Thm. 2.53]{amfupa00} we get  
$\mathscr{H}^{2}(\setG)
\geq \mathscr{H}^{2}(\pi(\setG))
=\mathscr{H}^{2}(\setA)
=\mathscr{L}^{2}(\setA)
>0$.  

\emph{Construction of $f$}: 
The mapping 
\begin{align}
f\colon\reals^3&\to\reals\\
\tp{(x_1\ x_2\ x_3)}&\mapsto x_3
\end{align}
is linear and one-to-one on $\setG$.
\end{exa}

The structure theorem in geometric measure theory \cite[Thm. 2.65]{amfupa00} implies that the  $2$-rectifiable set $\setG$ in Example \ref{exa.counter} is ``visible" from almost all directions, in the sense of the projection of $\setG$ onto a $2$-dimensional linear subspace in general position having positive Lebesgue measure. 
However, as just demonstrated, this does not prevent $\setG$ from being linearly compressible into $\reals$ in a one-to-one fashion.



For  $s$-rectifiable random vectors, $n\geq s$ is---in general---not necessary for successful recovery of $\rvecx$ and additional requirements on $\rvecx$ need to be imposed to get
converse statements of the form of what we would expect from classical compressed sensing theory. 
This leads us to the new concept of $s$-analytic measures and $s$-analytic random vectors.  We start with the definition of real analytic mappings. 

\begin{dfn}
We call 
\begin{enumerate}[(i)]
\item 
a function  $f\colon\reals^{k}\to \reals$ real analytic if, for each $\vecx\in\reals^k$, $f$ may be represented by a convergent power series in some neighborhood of $\vecx$;
\item 
a mapping  $f\colon\reals^{k}\to \reals^l$, $\vecx\mapsto \tp{(f_1(\vecx)\ \dots\ f_l(\vecx))}$ real analytic if each component $f_i$, $i=1,\dots,l$, is a real analytic function. 
\end{enumerate}
\end{dfn}

We are now ready to define the notion of $s$-analytic measures and $s$-analytic random vectors. 

\begin{dfn}\label{dfn:measureanalytic}
We call a Borel measure $\mu$ on $\reals^{m}$ $s$-analytic if for each $\setU\subseteq\reals^{m}$ with $\mu(\setU)>0$ we can find a real analytic mapping $h\colon\reals^s\to \reals^m$ of $s$-dimensional Jacobian $Jh\not\equiv 0$  and a set $\setA\subseteq\reals^s$ of positive Lebesgue measure such that $h(\setA)\subseteq\setU$. 
\end{dfn}

\begin{dfn}\label{dfnXana}
The random vector $\rvecx\in\reals^{m}$ is called  $s$-analytic if $\mu_\rvecx$ is $s$-analytic. 
\end{dfn}

It is instructive to compare $s$-analytic sets $\setU$ with $\mu(\setU)>0$ 
to the set $\setG$ in Example \ref{exa.counter}. 
Both $\setU$ and $\setG$ contain the image of a set with positive Lebesgue measure under a certain mapping. 
However, the mapping in Example \ref{exa.counter} is $C^1$, whereas the mapping in Definition \ref{dfn:measureanalytic} is real-analytic (with $Jh\not\equiv 0$). 
It turns out that real analyticity is strong enough to prevent $\setU$ from being mapped linearly in a one-to-one fashion into $\reals^t$ for $t<s$. 
Since this holds for every set $\setU$ with $\mu(\setU)>0$, $n\geq s$ is necessary for P2 to hold for $s$-analytic $\rvecx$.  
For if  there existed an $\varepsilon\in[0,1)$, an  $\matA\in \reals^{n\times m}$, and a decoder $g_\matA$ satisfying  \eqref{eq:decoder}, there would have to be 
a set $\setU\subseteq\reals^m$ with $\opP[\rvecx\in\setU]\geq 1-\varepsilon$ such that  $\matA$ is one-to-one on $\setU$, which is not possible for $n<s$ thanks to the analyticity of $\mu$. 

We are now ready to state our converse result for $s$-analytic random vectors. 

\begin{thm}\label{thm.converse}
Let $f\colon\reals^m\to \reals^n$ be a linear mapping,  $h\colon\reals^s\to \reals^m$  a real analytic mapping of $s$-dimensional Jacobian $Jh\not\equiv 0$,  and $\setA\subseteq\reals^s$ of positive Lebesgue measure. 
Suppose that $f$ is one-to-one on $h(\setA)$. 
Then $n\geq s$. 
\end{thm}

\begin{cor}\label{cor:analytic}
For $\rvecx\in\reals^m$ $s$-analytic, $n\geq s$ is  necessary for  P2 to hold.
\end{cor}
This result is, in fact, a strong converse as it  shows that for $n<s$ there is no pair $(\matA, g_\matA)$ such that  \eqref{eq:decoder} holds for 
$\varepsilon<1$. 
We close this section by establishing important properties of $s$-analytic measures, which will be used in the examples in the next section.  


\begin{lem}\label{lem:propalalytic}
Suppose that $\mu$ is $s$-analytic. Then,    
\begin{enumerate}[(i)]
\item  
$\mu$ is $t$-analytic for all $t\in\{1,\dots,s\}$;\label{propra2}
\item 
$\mu\ll\mathscr{H}^{s}$.\label{propra3} 
\end{enumerate}
\end{lem}

\section{Examples}

\begin{exa}\label{exrecana1}
Let $\rvecx\in\reals^{m}$ be as in Example \ref{exrecana1a}. Using the properties of the Gaussian distribution, a straightforward analysis reveals that $\rvecx$ is $s$-analytic. 
Furthermore, the $s$-rectifiable set $\setU$ in \eqref{eq:setS} satisfies $\opP[\rvecx\in\setU]=1$. 
Therefore, by \eqref{propra3} in Lemma \ref{lem:propalalytic}, $\rvecx$ is $s$-rectifiable. 
It follows from Corollary \ref{cor:rectifiable} that $n>s$ is sufficient for P1  to hold  and from Corollary \ref{cor:analytic} that $n\geq s$ is necessary for  P2 to hold. 
The information-theoretic limit we obtain here is best possible in the sense of classical compressed sensing where recovery thresholds suffer either from the square-root bottleneck or from a $\log(m)$-factor. We hasten to add, however, that we do not
specify decoders that achieve our threshold, rather we only
prove the existence of such decoders.
\end{exa}

The second example serves to demonstrate that a random vector's sparsity level in terms of the number of non-zero entries may differ vastly from its rectifiability and analyticity parameter. Specifically, we construct an $(r+t-1)$-rectifiable and $(r+t-1)$-analytic random vector with  sparsity level---in terms of the number of non-zero entries of the vector's realizations---$rt \gg (r+t-1)$. 
 
\begin{exa}\label{exrecana}
Let $\rvecx =\rveca\otimes\rvecb\in\reals^{kl}$, where $\rveca\in\reals^{k}$, $\rvecb\in\reals^{l}$,  and $\rveca$ and $\rvecb$ are statistically independent. 
Suppose that 
$\rveca$ has $r$ i.i.d. Gaussian entries  at positions drawn uniformly at random and all other entries equal  to zero and 
$\rvecb$ has $t$ i.i.d. Gaussian entries  at positions drawn uniformly at random and all other entries equal  to zero. 
Lemma \ref{auxlemma1} below  shows that $\rvecx$ is $(r+t-1)$-analytic. 
Furthermore, a straightforward analysis reveals that the $(r+t-1)$-rectifiable set 
\begin{align}
\setU&=\{\veca\otimes\vecb : \veca\in\tilde\setA_{r}, \vecb\in\setB_{t}\},
\end{align}
where
\begin{align}
\tilde\setA_{r}&=\{\veca\in\reals^k : \|\veca\|_0=r, a_\text{nz}=1\}\\
\setB_{t}&=\{\vecb\in\reals^l : \|\vecb\|_0=t\}
\end{align}
and $a_\text{nz}$ denotes the first non-zero entry of $\veca$, satisfies $\opP[\rvecx\in\setU]=1$. 
By  \eqref{propra3} in Lemma \ref{lem:propalalytic}, $\rvecx$ is $(r+t-1)$-rectifiable. 
It therefore follows from Corollary \ref{cor:rectifiable} that $n>(r+t-1)$ is sufficient for P1 to hold and from Corollary \ref{cor:analytic} that $n\geq (r+t-1)$ is necessary  for P2 to hold. 
Note that, for $r,t$ large, we have $(r+t-1)\ll rt$. 
What is interesting here is that the sparsity level of $\rvecx$---as quantified by the number of non-zero entries of the realizations of $\rvecx$---is $rt$, yet $r+t$ linear measurements suffice for recovery of $\rvecx$ with zero probability of error.
\end{exa}

\begin{lem}\label{auxlemma1}
Let $\rvecx =\rveca\otimes\rvecb\in\reals^{kl}$, where $\rveca\in\reals^{k}$ and $\rvecb\in\reals^{l}$ are random vectors such that $\mu_{\rveca}\times \mu_{\rvecb}\ll\mathscr{L}^{k+l}$. 
Then, $\rvecx$ is $(k+l-1)$-analytic.  
\end{lem}

\bibliographystyle{IEEEtran}
\bibliography{references}
\end{document}

%% file: math-macros.tex
\usepackage{amsmath}
\usepackage{amssymb}
\usepackage{amsfonts}
\usepackage{amsthm}
\usepackage{mathrsfs}
\usepackage{xspace}
\usepackage{bm}
\usepackage{fancyref}
\usepackage{textcomp}
\usepackage[Symbol]{upgreek}

\newcommand{\safemath}[2]{\newcommand{#1}{\ensuremath{#2}\xspace}}

\newcommand{\ssa}{\mathsf{a}}
\newcommand{\ssb}{\mathsf{b}}
\newcommand{\ssc}{\mathsf{c}}
\newcommand{\ssd}{\mathsf{d}}
\newcommand{\sse}{\mathsf{e}}
\newcommand{\ssf}{\mathsf{f}}
\newcommand{\ssg}{\mathsf{g}}
\newcommand{\ssh}{\mathsf{h}}
\newcommand{\ssi}{\mathsf{i}}
\newcommand{\ssj}{\mathsf{j}}
\newcommand{\ssk}{\mathsf{k}}
\newcommand{\ssl}{\mathsf{l}}
\newcommand{\ssm}{\mathsf{m}}
\newcommand{\ssn}{\mathsf{n}}
\newcommand{\sso}{\mathsf{o}}
\newcommand{\ssp}{\mathsf{p}}
\newcommand{\ssq}{\mathsf{q}}
\newcommand{\ssr}{\mathsf{r}}
\newcommand{\sss}{\mathsf{s}}
\newcommand{\sst}{\mathsf{t}}
\newcommand{\ssu}{\mathsf{u}}
\newcommand{\ssv}{\mathsf{v}}
\newcommand{\ssw}{\mathsf{w}}
\newcommand{\ssx}{\mathsf{x}}
\newcommand{\ssy}{\mathsf{y}}
\newcommand{\ssz}{\mathsf{z}}

\safemath{\bmsa}{\bm{\ssa}}
\safemath{\bmsb}{\bm{\ssb}}
\safemath{\bmsc}{\bm{\ssc}}
\safemath{\bmsd}{\bm{\ssd}}
\safemath{\bmse}{\bm{\sse}}
\safemath{\bmsf}{\bm{\ssf}}
\safemath{\bmsg}{\bm{\ssg}}
\safemath{\bmsh}{\bm{\ssh}}
\safemath{\bmsi}{\bm{\ssi}}
\safemath{\bmsj}{\bm{\ssj}}
\safemath{\bmsk}{\bm{\ssk}}
\safemath{\bmsl}{\bm{\ssl}}
\safemath{\bmsm}{\bm{\ssm}}
\safemath{\bmsn}{\bm{\ssn}}
\safemath{\bmso}{\bm{\sso}}
\safemath{\bmsp}{\bm{\ssp}}
\safemath{\bmsq}{\bm{\ssq}}
\safemath{\bmsr}{\bm{\ssr}}
\safemath{\bmss}{\bm{\sss}}
\safemath{\bmst}{\bm{\sst}}
\safemath{\bmsu}{\bm{\ssu}}
\safemath{\bmsv}{\bm{\ssv}}
\safemath{\bmsw}{\bm{\ssw}}
\safemath{\bmsx}{\bm{\ssx}}
\safemath{\bmsy}{\bm{\ssy}}
\safemath{\bmsz}{\bm{\ssz}}

\bmdefine{\bmualphad}{\upalpha}
\bmdefine{\bmubetad}{\upbeta}
\bmdefine{\bmuchid}{\upchi}
\bmdefine{\bmudeltad}{\updelta}
\bmdefine{\bmuepsilond}{\upepsilon}
\bmdefine{\bmuvarepsilond}{\upvarepsilon}
\bmdefine{\bmuetad}{\upeta}
\bmdefine{\bmugammad}{\upgamma}
\bmdefine{\bmuiotad}{\upiota}
\bmdefine{\bmukappad}{\upkappa}
\bmdefine{\bmulambdad}{\uplambda}
\bmdefine{\bmumud}{\upmu}
\bmdefine{\bmunud}{\upnu}
\bmdefine{\bmuomegad}{\upomega}
\bmdefine{\bmuphid}{\upphi}
\bmdefine{\bmuvarphid}{\upvarphi}
\bmdefine{\bmupid}{\uppi}
\bmdefine{\bmuvarpid}{\upvarpi}
\bmdefine{\bmupsid}{\uppsi}
\bmdefine{\bmurhod}{\uprho}
\bmdefine{\bmuvarrhod}{\upvarrho}
\bmdefine{\bmusigmad}{\upsigma}
\bmdefine{\bmuvarsigmad}{\upvarsigma}
\bmdefine{\bmutaud}{\uptau}
\bmdefine{\bmuthetad}{\uptheta}
\bmdefine{\bmuvarthetad}{\upvartheta}
\bmdefine{\bmuupsilond}{\upupsilon}
\bmdefine{\bmuxid}{\upxi}
\bmdefine{\bmuzetad}{\upzeta}

\safemath{\bmua}{\mathbf{a}}
\safemath{\bmub}{\mathbf{b}}
\safemath{\bmuc}{\mathbf{c}}
\safemath{\bmud}{\mathbf{d}}
\safemath{\bmue}{\mathbf{e}}
\safemath{\bmuf}{\mathbf{f}}
\safemath{\bmug}{\mathbf{g}}
\safemath{\bmuh}{\mathbf{h}}
\safemath{\bmui}{\mathbf{i}}
\safemath{\bmuj}{\mathbf{j}}
\safemath{\bmuk}{\mathbf{k}}
\safemath{\bmul}{\mathbf{l}}
\safemath{\bmum}{\mathbf{m}}
\safemath{\bmun}{\mathbf{n}}
\safemath{\bmuo}{\mathbf{o}}
\safemath{\bmup}{\mathbf{p}}
\safemath{\bmuq}{\mathbf{q}}
\safemath{\bmur}{\mathbf{r}}
\safemath{\bmus}{\mathbf{s}}
\safemath{\bmut}{\mathbf{t}}
\safemath{\bmuu}{\mathbf{u}}
\safemath{\bmuv}{\mathbf{v}}
\safemath{\bmuw}{\mathbf{w}}
\safemath{\bmux}{\mathbf{x}}
\safemath{\bmuy}{\mathbf{y}}
\safemath{\bmuz}{\mathbf{z}}

\safemath{\bmualpha}{\bmualphad}
\safemath{\bmubeta}{\bmubetad}
\safemath{\bmuchi}{\bumchid}
\safemath{\bmudelta}{\bmudeltad}
\safemath{\bmuepsilon}{\bmuepsilond}
\safemath{\bmuvarepsilon}{\bmuvarepsilond}
\safemath{\bmueta}{\bmuetad}
\safemath{\bmugamma}{\bmugammad}
\safemath{\bmuiota}{\bmuiotad}
\safemath{\bmukappa}{\bmukappad}
\safemath{\bmulambda}{\bmulambdad}
\safemath{\bmumu}{\bmumud}
\safemath{\bmunu}{\bmunud}
\safemath{\bmuomega}{\bmuomegad}
\safemath{\bmuphi}{\bmuphid}
\safemath{\bmuvarphi}{\bmuvarphid}
\safemath{\bmupi}{\bmupid}
\safemath{\bmuvarpi}{\bmuvarpid}
\safemath{\bmupsi}{\bmupsid}
\safemath{\bmurho}{\bmurhod}
\safemath{\bmuvarrho}{\bmuvarrhod}
\safemath{\bmusigma}{\bmusigmad}
\safemath{\bmuvarsigma}{\bmuvarsigmad}
\safemath{\bmutau}{\bmutaud}
\safemath{\bmutheta}{\bmuthetad}
\safemath{\bmuvartheta}{\bmuvarthetad}
\safemath{\bmuupsilon}{\bmuupsilond}
\safemath{\bmuxi}{\bmuxid}
\safemath{\bmuzeta}{\bmuzetad}

\bmdefine{\bmiad}{a}
\bmdefine{\bmibd}{b}
\bmdefine{\bmicd}{c}
\bmdefine{\bmidd}{d}
\bmdefine{\bmied}{e}
\bmdefine{\bmifd}{f}
\bmdefine{\bmigd}{g}
\bmdefine{\bmihd}{h}
\bmdefine{\bmiid}{i}
\bmdefine{\bmijd}{j}
\bmdefine{\bmikd}{k}
\bmdefine{\bmild}{l}
\bmdefine{\bmimd}{m}
\bmdefine{\bmind}{n}
\bmdefine{\bmiod}{o}
\bmdefine{\bmipd}{p}
\bmdefine{\bmiqd}{q}
\bmdefine{\bmird}{r}
\bmdefine{\bmisd}{s}
\bmdefine{\bmitd}{t}
\bmdefine{\bmiud}{u}
\bmdefine{\bmivd}{v}
\bmdefine{\bmiwd}{w}
\bmdefine{\bmixd}{x}
\bmdefine{\bmiyd}{y}
\bmdefine{\bmizd}{z}

\bmdefine{\bmialphad}{\alpha}
\bmdefine{\bmibetad}{\beta}
\bmdefine{\bmichid}{\chi}
\bmdefine{\bmideltad}{\delta}
\bmdefine{\bmiepsilond}{\epsilon}
\bmdefine{\bmivarepsilond}{\varepsilon}
\bmdefine{\bmietad}{\eta}
\bmdefine{\bmigammad}{\gamma}
\bmdefine{\bmiiotad}{\iota}
\bmdefine{\bmikappad}{\kappa}
\bmdefine{\bmivarkappad}{\varkappa}
\bmdefine{\bmilambdad}{\lambda}
\bmdefine{\bmimud}{\mu}
\bmdefine{\bminud}{\nu}
\bmdefine{\bmiomegad}{\omega}
\bmdefine{\bmiphid}{\phi}
\bmdefine{\bmivarphid}{\varphi}
\bmdefine{\bmipid}{\pi}
\bmdefine{\bmivarpid}{\varpi}
\bmdefine{\bmipsid}{\psi}
\bmdefine{\bmirhod}{\rho}
\bmdefine{\bmivarrhod}{\varrho}
\bmdefine{\bmisigmad}{\sigma}
\bmdefine{\bmivarsigmad}{\varsigma}
\bmdefine{\bmitaud}{\tau}
\bmdefine{\bmithetad}{\theta}
\bmdefine{\bmivarthetad}{\vartheta}
\bmdefine{\bmiupsilond}{\upsilon}
\bmdefine{\bmixid}{\xi}
\bmdefine{\bmizetad}{\zeta}

\safemath{\bmia}{\bmiad}
\safemath{\bmib}{\bmibd}
\safemath{\bmic}{\bmicd}
\safemath{\bmid}{\bmidd}
\safemath{\bmie}{\bmied}
\safemath{\bmif}{\bmifd}
\safemath{\bmig}{\bmigd}
\safemath{\bmih}{\bmihd}
\safemath{\bmii}{\bmiid}
\safemath{\bmij}{\bmijd}
\safemath{\bmik}{\bmikd}
\safemath{\bmil}{\bmild}
\safemath{\bmim}{\bmimd}
\safemath{\bmin}{\bmind}
\safemath{\bmio}{\bmiod}
\safemath{\bmip}{\bmipd}
\safemath{\bmiq}{\bmiqd}
\safemath{\bmir}{\bmird}
\safemath{\bmis}{\bmisd}
\safemath{\bmit}{\bmitd}
\safemath{\bmiu}{\bmiud}
\safemath{\bmiv}{\bmivd}
\safemath{\bmiw}{\bmiwd}
\safemath{\bmix}{\bmixd}
\safemath{\bmiy}{\bmiyd}
\safemath{\bmiz}{\bmizd}

\safemath{\bmialpha}{\bmialphad}
\safemath{\bmibeta}{\bmibetad}
\safemath{\bmichi}{\bmichid}
\safemath{\bmidelta}{\bmideltad}
\safemath{\bmiepsilon}{\bmiepsilond}
\safemath{\bmivarepsilon}{\bmivarepsilond}
\safemath{\bmieta}{\bmietad}
\safemath{\bmigamma}{\bmigammad}
\safemath{\bmiiota}{\bmiiotad}
\safemath{\bmikappa}{\bmikappad}
\safemath{\bmivarkappa}{\bmivarkappad}
\safemath{\bmilambda}{\bmilambdad}
\safemath{\bmimu}{\bmimud}
\safemath{\bminu}{\bminud}
\safemath{\bmiomega}{\bmiomegad}
\safemath{\bmiphi}{\bmiphid}
\safemath{\bmivarphi}{\bmivarphid}
\safemath{\bmipi}{\bmipid}
\safemath{\bmivarpi}{\bmivarpid}
\safemath{\bmipsi}{\bmipsid}
\safemath{\bmirho}{\bmirhod}
\safemath{\bmivarrho}{\bmivarrhod}
\safemath{\bmisigma}{\bmisigmad}
\safemath{\bmivarsigma}{\bmivarsigmad}
\safemath{\bmitau}{\bmitaud}
\safemath{\bmitheta}{\bmithetad}
\safemath{\bmivartheta}{\bmivarthetad}
\safemath{\bmiupsilon}{\bmiupsilond}
\safemath{\bmixi}{\bmixid}
\safemath{\bmizeta}{\bmizetad}

\bmdefine{\bmuDeltad}{\Updelta}
\bmdefine{\bmuGammad}{\Upgamma}
\bmdefine{\bmuLambdad}{\Uplambda}
\bmdefine{\bmuOmegad}{\Upomega}
\bmdefine{\bmuPhid}{\Upphi}
\bmdefine{\bmuPid}{\Uppi}
\bmdefine{\bmuPsid}{\Uppsi}
\bmdefine{\bmuSigmad}{\Upsigma}
\bmdefine{\bmuThetad}{\Uptheta}
\bmdefine{\bmuUpsilond}{\Upupsilon}
\bmdefine{\bmuXid}{\Upxi}

\safemath{\bmuA}{\mathbf{A}}
\safemath{\bmuB}{\mathbf{B}}
\safemath{\bmuC}{\mathbf{C}}
\safemath{\bmuD}{\mathbf{D}}
\safemath{\bmuE}{\mathbf{E}}
\safemath{\bmuF}{\mathbf{F}}
\safemath{\bmuG}{\mathbf{G}}
\safemath{\bmuH}{\mathbf{H}}
\safemath{\bmuI}{\mathbf{I}}
\safemath{\bmuJ}{\mathbf{J}}
\safemath{\bmuK}{\mathbf{K}}
\safemath{\bmuL}{\mathbf{L}}
\safemath{\bmuM}{\mathbf{M}}
\safemath{\bmuN}{\mathbf{N}}
\safemath{\bmuO}{\mathbf{O}}
\safemath{\bmuP}{\mathbf{P}}
\safemath{\bmuQ}{\mathbf{Q}}
\safemath{\bmuR}{\mathbf{R}}
\safemath{\bmuS}{\mathbf{S}}
\safemath{\bmuT}{\mathbf{T}}
\safemath{\bmuU}{\mathbf{U}}
\safemath{\bmuV}{\mathbf{V}}
\safemath{\bmuW}{\mathbf{W}}
\safemath{\bmuX}{\mathbf{X}}
\safemath{\bmuY}{\mathbf{Y}}
\safemath{\bmuZ}{\mathbf{Z}}

\safemath{\bmuZero}{\mathbf{0}}
\safemath{\bmuOne}{\mathbf{1}}

\safemath{\bmuDelta}{\bmuDeltad}
\safemath{\bmuGamma}{\bmuGammad}
\safemath{\bmuLambda}{\bmuLambdad}
\safemath{\bmuOmega}{\bmuOmegad}
\safemath{\bmuPhi}{\bmuPhid}
\safemath{\bmuPi}{\bmuPid}
\safemath{\bmuPsi}{\bmuPsid}
\safemath{\bmuSigma}{\bmuSigmad}
\safemath{\bmuTheta}{\bmuThetad}
\safemath{\bmuUpsilon}{\bmuUpsilond}
\safemath{\bmuXi}{\bmuXid}

\bmdefine{\bmiAd}{A}
\bmdefine{\bmiBd}{B}
\bmdefine{\bmiCd}{C}
\bmdefine{\bmiDd}{D}
\bmdefine{\bmiEd}{E}
\bmdefine{\bmiFd}{F}
\bmdefine{\bmiGd}{G}
\bmdefine{\bmiHd}{H}
\bmdefine{\bmiId}{I}
\bmdefine{\bmiJd}{J}
\bmdefine{\bmiKd}{K}
\bmdefine{\bmiLd}{L}
\bmdefine{\bmiMd}{M}
\bmdefine{\bmiOd}{N}
\bmdefine{\bmiPd}{O}
\bmdefine{\bmiQd}{P}
\bmdefine{\bmiRd}{R}
\bmdefine{\bmiSd}{S}
\bmdefine{\bmiTd}{T}
\bmdefine{\bmiUd}{U}
\bmdefine{\bmiVd}{V}
\bmdefine{\bmiWd}{W}
\bmdefine{\bmiXd}{X}
\bmdefine{\bmiYd}{Y}
\bmdefine{\bmiZd}{Z}

\bmdefine{\bmiDeltad}{\Delta}
\bmdefine{\bmiGammad}{\Gamma}
\bmdefine{\bmiLambdad}{\Lambda}
\bmdefine{\bmiOmegad}{\Omega}
\bmdefine{\bmiPhid}{\Phi}
\bmdefine{\bmiPid}{\Pi}
\bmdefine{\bmiPsid}{\Psi}
\bmdefine{\bmiSigmad}{\Sigma}
\bmdefine{\bmiThetad}{\Theta}
\bmdefine{\bmiUpsilond}{\Upsilon}
\bmdefine{\bmiXid}{\Xi}

\safemath{\bmiA}{\bmiAd}
\safemath{\bmiB}{\bmiBd}
\safemath{\bmiC}{\bmiCd}
\safemath{\bmiD}{\bmiDd}
\safemath{\bmiE}{\bmiEd}
\safemath{\bmiF}{\bmiFd}
\safemath{\bmiG}{\bmiGd}
\safemath{\bmiH}{\bmiHd}
\safemath{\bmiI}{\bmiId}
\safemath{\bmiJ}{\bmiJd}
\safemath{\bmiK}{\bmiKd}
\safemath{\bmiL}{\bmiLd}
\safemath{\bmiM}{\bmiMd}
\safemath{\bmiN}{\bmiNd}
\safemath{\bmiO}{\bmiOd}
\safemath{\bmiP}{\bmiPd}
\safemath{\bmiQ}{\bmiQd}
\safemath{\bmiR}{\bmiRd}
\safemath{\bmiS}{\bmiSd}
\safemath{\bmiT}{\bmiTd}
\safemath{\bmiU}{\bmiUd}
\safemath{\bmiV}{\bmiVd}
\safemath{\bmiW}{\bmiWd}
\safemath{\bmiX}{\bmiXd}
\safemath{\bmiY}{\bmiYd}
\safemath{\bmiZ}{\bmiZd}

\safemath{\bmiDelta}{\bmiDeltad}
\safemath{\bmiGamma}{\bmiGammad}
\safemath{\bmiLambda}{\bmiLambdad}
\safemath{\bmiOmega}{\bmiOmegad}
\safemath{\bmiPhi}{\bmiPhid}
\safemath{\bmiPi}{\bmiPid}
\safemath{\bmiPsi}{\bmiPsid}
\safemath{\bmiSigma}{\bmiSigmad}
\safemath{\bmiTheta}{\bmiThetad}
\safemath{\bmiUpsilon}{\bmiUpsilond}
\safemath{\bmiXi}{\bmiXid}

\safemath{\evA}{\mathcal{A}}
\safemath{\evB}{\mathcal{B}}
\safemath{\evC}{\mathcal{C}}
\safemath{\evD}{\mathcal{D}}
\safemath{\evE}{\mathcal{E}}
\safemath{\evF}{\mathcal{F}}
\safemath{\evG}{\mathcal{G}}
\safemath{\evH}{\mathcal{H}}
\safemath{\evI}{\mathcal{I}}
\safemath{\evJ}{\mathcal{J}}
\safemath{\evK}{\mathcal{K}}
\safemath{\evL}{\mathcal{L}}
\safemath{\evM}{\mathcal{M}}
\safemath{\evN}{\mathcal{N}}
\safemath{\evO}{\mathcal{O}}
\safemath{\evP}{\mathcal{P}}
\safemath{\evQ}{\mathcal{Q}}
\safemath{\evR}{\mathcal{R}}
\safemath{\evS}{\mathcal{S}}
\safemath{\evT}{\mathcal{T}}
\safemath{\evU}{\mathcal{U}}
\safemath{\evV}{\mathcal{V}}
\safemath{\evW}{\mathcal{W}}
\safemath{\evX}{\mathcal{X}}
\safemath{\evY}{\mathcal{Y}}
\safemath{\evZ}{\mathcal{Z}}

\safemath{\setA}{\mathcal{A}}
\safemath{\setB}{\mathcal{B}}
\safemath{\setC}{\mathcal{C}}
\safemath{\setD}{\mathcal{D}}
\safemath{\setE}{\mathcal{E}}
\safemath{\setF}{\mathcal{F}}
\safemath{\setG}{\mathcal{G}}
\safemath{\setH}{\mathcal{H}}
\safemath{\setI}{\mathcal{I}}
\safemath{\setJ}{\mathcal{J}}
\safemath{\setK}{\mathcal{K}}
\safemath{\setL}{\mathcal{L}}
\safemath{\setM}{\mathcal{M}}
\safemath{\setN}{\mathcal{N}}
\safemath{\setO}{\mathcal{O}}
\safemath{\setP}{\mathcal{P}}
\safemath{\setQ}{\mathcal{Q}}
\safemath{\setR}{\mathcal{R}}
\safemath{\setS}{\mathcal{S}}
\safemath{\setT}{\mathcal{T}}
\safemath{\setU}{\mathcal{U}}
\safemath{\setV}{\mathcal{V}}
\safemath{\setW}{\mathcal{W}}
\safemath{\setX}{\mathcal{X}}
\safemath{\setY}{\mathcal{Y}}
\safemath{\setZ}{\mathcal{Z}}
\safemath{\emptySet}{\varnothing}

\safemath{\colA}{\mathscr{A}}
\safemath{\colB}{\mathscr{B}}
\safemath{\colC}{\mathscr{C}}
\safemath{\colD}{\mathscr{D}}
\safemath{\colE}{\mathscr{E}}
\safemath{\colF}{\mathscr{F}}
\safemath{\colG}{\mathscr{G}}
\safemath{\colH}{\mathscr{H}}
\safemath{\colI}{\mathscr{I}}
\safemath{\colJ}{\mathscr{J}}
\safemath{\colK}{\mathscr{K}}
\safemath{\colL}{\mathscr{L}}
\safemath{\colM}{\mathscr{M}}
\safemath{\colN}{\mathscr{N}}
\safemath{\colO}{\mathscr{O}}
\safemath{\colP}{\mathscr{P}}
\safemath{\colQ}{\mathscr{Q}}
\safemath{\colR}{\mathscr{R}}
\safemath{\colS}{\mathscr{S}}
\safemath{\colT}{\mathscr{T}}
\safemath{\colU}{\mathscr{U}}
\safemath{\colV}{\mathscr{V}}
\safemath{\colW}{\mathscr{W}}
\safemath{\colX}{\mathscr{X}}
\safemath{\colY}{\mathscr{Y}}
\safemath{\colZ}{\mathscr{Z}}

\safemath{\opA}{\mathbb{A}}
\safemath{\opB}{\mathbb{B}}
\safemath{\opC}{\mathbb{C}}
\safemath{\opD}{\mathbb{D}}
\safemath{\opE}{\mathbb{E}}
\safemath{\opF}{\mathbb{F}}
\safemath{\opG}{\mathbb{G}}
\safemath{\opH}{\mathbb{H}}
\safemath{\opI}{\mathbb{I}}
\safemath{\opJ}{\mathbb{J}}
\safemath{\opK}{\mathbb{K}}
\safemath{\opL}{\mathbb{L}}
\safemath{\opM}{\mathbb{M}}
\safemath{\opN}{\mathbb{N}}
\safemath{\opO}{\mathbb{O}}
\safemath{\opP}{\mathbb{P}}
\safemath{\opQ}{\mathbb{Q}}
\safemath{\opR}{\mathbb{R}}
\safemath{\opS}{\mathbb{S}}
\safemath{\opT}{\mathbb{T}}
\safemath{\opU}{\mathbb{U}}
\safemath{\opV}{\mathbb{V}}
\safemath{\opW}{\mathbb{W}}
\safemath{\opX}{\mathbb{X}}
\safemath{\opY}{\mathbb{Y}}
\safemath{\opZ}{\mathbb{Z}}
\safemath{\opZero}{\mathbb{O}}
\safemath{\identityop}{\opI}

\safemath{\sca}{a}
\safemath{\scb}{b}
\safemath{\scc}{c}
\safemath{\scd}{d}
\safemath{\sce}{e}
\safemath{\scf}{f}
\safemath{\scg}{g}
\safemath{\sch}{h}
\safemath{\sci}{i}
\safemath{\scj}{j}
\safemath{\sck}{k}
\safemath{\scl}{l}
\safemath{\scm}{m}
\safemath{\scn}{n}
\safemath{\sco}{o}
\safemath{\scp}{p}
\safemath{\scq}{q}
\safemath{\scr}{r}
\safemath{\scs}{s}
\safemath{\sct}{t}
\safemath{\scu}{u}
\safemath{\scv}{v}
\safemath{\scw}{w}
\safemath{\scx}{x}
\safemath{\scy}{y}
\safemath{\scz}{z}

\safemath{\scA}{A}
\safemath{\scB}{B}
\safemath{\scC}{C}
\safemath{\scD}{D}
\safemath{\scE}{E}
\safemath{\scF}{F}
\safemath{\scG}{G}
\safemath{\scH}{H}
\safemath{\scI}{I}
\safemath{\scJ}{J}
\safemath{\scK}{K}
\safemath{\scL}{L}
\safemath{\scM}{M}
\safemath{\scN}{N}
\safemath{\scO}{O}
\safemath{\scP}{P}
\safemath{\scQ}{Q}
\safemath{\scR}{R}
\safemath{\scS}{S}
\safemath{\scT}{T}
\safemath{\scU}{U}
\safemath{\scV}{V}
\safemath{\scW}{W}
\safemath{\scX}{X}
\safemath{\scY}{Y}
\safemath{\scZ}{Z}

\safemath{\scalpha}{\alpha}
\safemath{\scbeta}{\beta}
\safemath{\scchi}{\chi}
\safemath{\scdelta}{\delta}
\safemath{\scepsilon}{\epsilon}
\safemath{\scvarepsilon}{\varepsilon}
\safemath{\sceta}{\eta}
\safemath{\scgamma}{\gamma}
\safemath{\sciota}{\iota}
\safemath{\sckappa}{\kappa}
\safemath{\scvarkappa}{\varkappa}
\safemath{\sclambda}{\lambda}
\safemath{\scmu}{\mu}
\safemath{\scnu}{\nu}
\safemath{\scomega}{\omega}
\safemath{\scphi}{\phi}
\safemath{\scvarphi}{\varphi}
\safemath{\scpi}{\pi}
\safemath{\scvarpi}{\varpi}
\safemath{\scpsi}{\psi}
\safemath{\scrho}{\rho}
\safemath{\scvarrho}{\varrho}
\safemath{\scsigma}{\sigma}
\safemath{\scvarsigma}{\varsigma}
\safemath{\sctau}{\tau}
\safemath{\sctheta}{\theta}
\safemath{\scvartheta}{\vartheta}
\safemath{\scupsilon}{\upsilon}
\safemath{\scxi}{\xi}
\safemath{\sczeta}{\zeta}

\safemath{\veca}{{\boldsymbol{a}}}
\safemath{\vecb}{{\boldsymbol{b}}}
\safemath{\vecc}{{\boldsymbol{c}}}
\safemath{\vecd}{{\boldsymbol{d}}}
\safemath{\vece}{{\boldsymbol{e}}}
\safemath{\vecf}{{\boldsymbol{f}}}
\safemath{\vecg}{{\boldsymbol{g}}}
\safemath{\vech}{{\boldsymbol{h}}}
\safemath{\veci}{{\boldsymbol{i}}}
\safemath{\vecj}{{\boldsymbol{j}}}
\safemath{\veck}{{\boldsymbol{k}}}
\safemath{\vecl}{{\boldsymbol{l}}}
\safemath{\vecm}{{\boldsymbol{m}}}
\safemath{\vecn}{{\boldsymbol{n}}}
\safemath{\veco}{{\boldsymbol{o}}}
\safemath{\vecp}{{\boldsymbol{p}}}
\safemath{\vecq}{{\boldsymbol{q}}}
\safemath{\vecr}{{\boldsymbol{r}}}
\safemath{\vecs}{{\boldsymbol{s}}}
\safemath{\vect}{{\boldsymbol{t}}}
\safemath{\vecu}{{\boldsymbol{u}}}
\safemath{\vecv}{{\boldsymbol{v}}}
\safemath{\vecw}{{\boldsymbol{w}}}
\safemath{\vecx}{{\boldsymbol{x}}}
\safemath{\vecy}{{\boldsymbol{y}}}
\safemath{\vecz}{{\boldsymbol{z}}}
\safemath{\veczero}{{\boldsymbol{0}}}
\safemath{\vecone}{{\boldsymbol{1}}}

\safemath{\vecalpha}{\upalpha}
\safemath{\vecbeta}{\upbeta}
\safemath{\vecchi}{\upchi}
\safemath{\vecdelta}{\updelta}
\safemath{\vecepsilon}{\upepsilon}
\safemath{\vecvarepsilon}{\upvarepsilon}
\safemath{\veceta}{\upeta}
\safemath{\vecgamma}{\upgamma}
\safemath{\veciota}{\upiota}
\safemath{\veckappa}{\upkappa}
\safemath{\veclambda}{\uplambda}
\safemath{\vecmu}{\text{\textmu}}
\safemath{\vecnu}{\upnu}
\safemath{\vecomega}{\upomega}
\safemath{\vecphi}{\upphi}
\safemath{\vecvarphi}{\upvarphi}
\safemath{\vecpi}{\uppi}
\safemath{\vecvarpi}{\upvarpi}
\safemath{\vecpsi}{\uppsi}
\safemath{\vecrho}{\uprho}
\safemath{\vecvarrho}{\upvarrho}
\safemath{\vecsigma}{\upsigma}
\safemath{\vecvarsigma}{\upvarsigma}
\safemath{\vectau}{\uptau}
\safemath{\vectheta}{\uptheta}
\safemath{\vecvartheta}{\upvartheta}
\safemath{\vecupsilon}{\upupsilon}
\safemath{\vecxi}{\upxi}
\safemath{\veczeta}{\upzeta}

\safemath{\vecac}{a}
\safemath{\vecbc}{b}
\safemath{\veccc}{c}
\safemath{\vecdc}{d}
\safemath{\vecec}{e}
\safemath{\vecfc}{f}
\safemath{\vecgc}{g}
\safemath{\vechc}{h}
\safemath{\vecic}{i}
\safemath{\vecjc}{j}
\safemath{\veckc}{k}
\safemath{\veclc}{l}
\safemath{\vecmc}{m}
\safemath{\vecnc}{n}
\safemath{\vecoc}{o}
\safemath{\vecpc}{p}
\safemath{\vecqc}{q}
\safemath{\vecrc}{r}
\safemath{\vecsc}{s}
\safemath{\vectc}{t}
\safemath{\vecuc}{u}
\safemath{\vecvc}{v}
\safemath{\vecwc}{w}
\safemath{\vecxc}{x}
\safemath{\vecyc}{y}
\safemath{\veczc}{z}

\safemath{\matA}{{\boldsymbol{A}}}
\safemath{\matB}{{\boldsymbol{B}}}
\safemath{\matC}{{\boldsymbol{C}}}
\safemath{\matD}{{\boldsymbol{D}}}
\safemath{\matE}{{\boldsymbol{E}}}
\safemath{\matF}{{\boldsymbol{F}}}
\safemath{\matG}{{\boldsymbol{G}}}
\safemath{\matH}{{\boldsymbol{H}}}
\safemath{\matI}{{\boldsymbol{I}}}
\safemath{\matJ}{{\boldsymbol{J}}}
\safemath{\matK}{{\boldsymbol{K}}}
\safemath{\matL}{{\boldsymbol{L}}}
\safemath{\matM}{{\boldsymbol{M}}}
\safemath{\matN}{{\boldsymbol{N}}}
\safemath{\matO}{{\boldsymbol{O}}}
\safemath{\matP}{{\boldsymbol{P}}}
\safemath{\matQ}{{\boldsymbol{Q}}}
\safemath{\matR}{{\boldsymbol{R}}}
\safemath{\matS}{{\boldsymbol{S}}}
\safemath{\matT}{{\boldsymbol{T}}}
\safemath{\matU}{{\boldsymbol{U}}}
\safemath{\matV}{{\boldsymbol{V}}}
\safemath{\matW}{{\boldsymbol{W}}}
\safemath{\matX}{{\boldsymbol{X}}}
\safemath{\matY}{{\boldsymbol{Y}}}
\safemath{\matZ}{{\boldsymbol{Z}}}
\safemath{\matzero}{{\boldsymbol{0}}}

\safemath{\matDelta}{\Updelta}
\safemath{\matGamma}{\Upgammma}
\safemath{\matLambda}{\Uplambda}
\safemath{\matOmega}{\Upomega}
\safemath{\matPhi}{\Upphi}
\safemath{\matPi}{\Uppi}
\safemath{\matPsi}{\Uppsi}
\safemath{\matSigma}{\Upsigma}
\safemath{\matTheta}{\Uptheta}
\safemath{\matUpsilon}{\Upupsilon}
\safemath{\matXi}{\Upxi}

\safemath{\matidentity}{\matI}
\safemath{\vecunit}{\vece} 
\safemath{\matone}{\matO}

\safemath{\matAc}{a}
\safemath{\matBc}{b}
\safemath{\matCc}{c}
\safemath{\matDc}{d}
\safemath{\matEc}{e}
\safemath{\matFc}{f}
\safemath{\matGc}{g}
\safemath{\matHc}{h}
\safemath{\matIc}{i}
\safemath{\matJc}{j}
\safemath{\matKc}{k}
\safemath{\matLc}{l}
\safemath{\matMc}{m}
\safemath{\matNc}{n}
\safemath{\matOc}{o}
\safemath{\matPc}{p}
\safemath{\matQc}{q}
\safemath{\matRc}{r}
\safemath{\matSc}{s}
\safemath{\matTc}{t}
\safemath{\matUc}{u}
\safemath{\matVc}{v}
\safemath{\matWc}{w}
\safemath{\matXc}{x}
\safemath{\matYc}{y}
\safemath{\matZc}{z}

\safemath{\rnda}{\mathsf{a}}
\safemath{\rndb}{\mathsf{b}}
\safemath{\rndc}{\mathsf{c}}
\safemath{\rndd}{\mathsf{d}}
\safemath{\rnde}{\mathsf{e}}
\safemath{\rndf}{\mathsf{f}}
\safemath{\rndg}{\mathsf{g}}
\safemath{\rndh}{\mathsf{h}}
\safemath{\rndi}{\mathsf{i}}
\safemath{\rndj}{\mathsf{j}}
\safemath{\rndk}{\mathsf{k}}
\safemath{\rndl}{\mathsf{l}}
\safemath{\rndm}{\mathsf{m}}
\safemath{\rndn}{\mathsf{n}}
\safemath{\rndo}{\mathsf{o}}
\safemath{\rndp}{\mathsf{p}}
\safemath{\rndq}{\mathsf{q}}
\safemath{\rndr}{\mathsf{r}}
\safemath{\rnds}{\mathsf{s}}
\safemath{\rndt}{\mathsf{t}}
\safemath{\rndu}{\mathsf{u}}
\safemath{\rndv}{\mathsf{v}}
\safemath{\rndw}{\mathsf{w}}
\safemath{\rndx}{\mathsf{x}}
\safemath{\rndy}{\mathsf{y}}
\safemath{\rndz}{\mathsf{z}}

\safemath{\rndA}{\bmiA}
\safemath{\rndB}{\bmiB}
\safemath{\rndC}{\bmiC}
\safemath{\rndD}{\bmiD}
\safemath{\rndE}{\bmiE}
\safemath{\rndF}{\bmiF}
\safemath{\rndG}{\bmiG}
\safemath{\rndH}{\bmiH}
\safemath{\rndI}{\bmiI}
\safemath{\rndJ}{\bmiJ}
\safemath{\rndK}{\bmiK}
\safemath{\rndL}{\bmiL}
\safemath{\rndM}{\bmiM}
\safemath{\rndN}{\bmiN}
\safemath{\rndO}{\bmiO}
\safemath{\rndP}{\bmiP}
\safemath{\rndQ}{\bmiQ}
\safemath{\rndR}{\bmiR}
\safemath{\rndS}{\bmiS}
\safemath{\rndT}{\bmiT}
\safemath{\rndU}{\bmiU}
\safemath{\rndV}{\bmiV}
\safemath{\rndW}{\bmiW}
\safemath{\rndX}{\bmiX}
\safemath{\rndY}{\bmiY}
\safemath{\rndZ}{\bmiZ}

\safemath{\rndalpha}{\bmialpha}
\safemath{\rndbeta}{\bmibeta}
\safemath{\rndchi}{\bmichi}
\safemath{\rnddelta}{\bmidelta}
\safemath{\rndepsilon}{\bmiepsilon}
\safemath{\rndvarepsilon}{\bmivarepsilon}
\safemath{\rndeta}{\bmieta}
\safemath{\rndgamma}{\bmigamma}
\safemath{\rndiota}{\bmiiota}
\safemath{\rndkappa}{\bmikappa}
\safemath{\rndlambda}{\bmilambda}
\safemath{\rndmu}{\bmimu}
\safemath{\rndnu}{\bminu}
\safemath{\rndomega}{\bmiomega}
\safemath{\rndphi}{\bmiphi}
\safemath{\rndvarphi}{\bmivarphi}
\safemath{\rndpi}{\bmipi}
\safemath{\rndvarpi}{\bmivarpi}
\safemath{\rndpsi}{\bmipsi}
\safemath{\rndrho}{\bmirho}
\safemath{\rndvarrho}{\bmivarrho}
\safemath{\rndsigma}{\bmisigma}
\safemath{\rndvarsigma}{\bmivarsigma}
\safemath{\rndtau}{\bmitau}
\safemath{\rndtheta}{\bmitheta}
\safemath{\rndvartheta}{\bmivartheta}
\safemath{\rndupsilon}{\bmiupsilon}
\safemath{\rndxi}{\bmixi}
\safemath{\rndzeta}{\bmizeta}

\safemath{\rveca}{{\boldsymbol{\mathsf{a}}}}
\safemath{\rvecb}{{\boldsymbol{\mathsf{b}}}}
\safemath{\rvecc}{{\boldsymbol{\mathsf{c}}}}
\safemath{\rvecd}{{\boldsymbol{\mathsf{d}}}}
\safemath{\rvece}{{\boldsymbol{\mathsf{e}}}}
\safemath{\rvecf}{{\boldsymbol{\mathsf{f}}}}
\safemath{\rvecg}{{\boldsymbol{\mathsf{g}}}}
\safemath{\rvech}{{\boldsymbol{\mathsf{h}}}}
\safemath{\rveci}{{\boldsymbol{\mathsf{i}}}}
\safemath{\rvecj}{{\boldsymbol{\mathsf{j}}}}
\safemath{\rveck}{{\boldsymbol{\mathsf{k}}}}
\safemath{\rvecl}{{\boldsymbol{\mathsf{l}}}}
\safemath{\rvecm}{{\boldsymbol{\mathsf{m}}}}
\safemath{\rvecn}{{\boldsymbol{\mathsf{n}}}}
\safemath{\rveco}{{\boldsymbol{\mathsf{o}}}}
\safemath{\rvecp}{{\boldsymbol{\mathsf{p}}}}
\safemath{\rvecq}{{\boldsymbol{\mathsf{q}}}}
\safemath{\rvecr}{{\boldsymbol{\mathsf{r}}}}
\safemath{\rvecs}{{\boldsymbol{\mathsf{s}}}}
\safemath{\rvect}{{\boldsymbol{\mathsf{t}}}}
\safemath{\rvecu}{{\boldsymbol{\mathsf{u}}}}
\safemath{\rvecv}{{\boldsymbol{\mathsf{v}}}}
\safemath{\rvecw}{{\boldsymbol{\mathsf{w}}}}
\safemath{\rvecx}{{\boldsymbol{\mathsf{x}}}}
\safemath{\rvecy}{{\boldsymbol{\mathsf{y}}}}
\safemath{\rvecz}{{\boldsymbol{\mathsf{z}}}}

\safemath{\rvecalpha}{\bmualpha}
\safemath{\rvecbeta}{\bmubeta}
\safemath{\rvecchi}{\bmuchi}
\safemath{\rvecdelta}{\bmudelta}
\safemath{\rvecepsilon}{\bmuepsilon}
\safemath{\rvecvarepsilon}{\bmuvarepsilon}
\safemath{\rveceta}{\bmueta}
\safemath{\rvecgamma}{\bmugamma}
\safemath{\rveciota}{\bmuiota}
\safemath{\rveckappa}{\bmukappa}
\safemath{\rveclambda}{\bmulambda}
\safemath{\rvecmu}{\bmumu}
\safemath{\rvecnu}{\bmunu}
\safemath{\rvecomega}{\bmuomega}
\safemath{\rvecphi}{\bmuphi}
\safemath{\rvecvarphi}{\bmuvarphi}
\safemath{\rvecpi}{\bmupi}
\safemath{\rvecvarpi}{\bmuvarpi}
\safemath{\rvecpsi}{\bmupsi}
\safemath{\rvecrho}{\bmurho}
\safemath{\rvecvarrho}{\bmuvarrho}
\safemath{\rvecsigma}{\bmusigma}
\safemath{\rvecvarsigma}{\bmuvarsigma}
\safemath{\rvectau}{\bmutau}
\safemath{\rvectheta}{\bmutheta}
\safemath{\rvecvartheta}{\bmuvartheta}
\safemath{\rvecupsilon}{\bmuupsilon}
\safemath{\rvecxi}{\bmuxi}
\safemath{\rveczeta}{\bmuzeta}

\safemath{\rvecac}{\rnda}
\safemath{\rvecbc}{\rndb}
\safemath{\rveccc}{\rndc}
\safemath{\rvecdc}{\rndd}
\safemath{\rvecec}{\rnde}
\safemath{\rvecfc}{\rndf}
\safemath{\rvecgc}{\rndg}
\safemath{\rvechc}{\rndh}
\safemath{\rvecic}{\rndi}
\safemath{\rvecjc}{\rndj}
\safemath{\rveckc}{\rndk}
\safemath{\rveclc}{\rndl}
\safemath{\rvecmc}{\rndm}
\safemath{\rvecnc}{\rndn}
\safemath{\rvecoc}{\rndo}
\safemath{\rvecpc}{\rndp}
\safemath{\rvecqc}{\rndq}
\safemath{\rvecrc}{\rndr}
\safemath{\rvecsc}{\rnds}
\safemath{\rvectc}{\rndt}
\safemath{\rvecuc}{\rndu}
\safemath{\rvecvc}{\rndv}
\safemath{\rvecwc}{\rndw}
\safemath{\rvecxc}{\rndx}
\safemath{\rvecyc}{\rndy}
\safemath{\rveczc}{\rndz}

\safemath{\rmatA}{{\boldsymbol{\mathsf{A}}}}
\safemath{\rmatB}{{\boldsymbol{\mathsf{B}}}}
\safemath{\rmatC}{{\boldsymbol{\mathsf{C}}}}
\safemath{\rmatD}{{\boldsymbol{\mathsf{D}}}}
\safemath{\rmatE}{{\boldsymbol{\mathsf{E}}}}
\safemath{\rmatF}{{\boldsymbol{\mathsf{F}}}}
\safemath{\rmatG}{{\boldsymbol{\mathsf{G}}}}
\safemath{\rmatH}{{\boldsymbol{\mathsf{H}}}}
\safemath{\rmatI}{{\boldsymbol{\mathsf{I}}}}
\safemath{\rmatJ}{{\boldsymbol{\mathsf{J}}}}
\safemath{\rmatK}{{\boldsymbol{\mathsf{K}}}}
\safemath{\rmatL}{{\boldsymbol{\mathsf{L}}}}
\safemath{\rmatM}{{\boldsymbol{\mathsf{M}}}}
\safemath{\rmatN}{{\boldsymbol{\mathsf{N}}}}
\safemath{\rmatO}{{\boldsymbol{\mathsf{O}}}}
\safemath{\rmatP}{{\boldsymbol{\mathsf{P}}}}
\safemath{\rmatQ}{{\boldsymbol{\mathsf{Q}}}}
\safemath{\rmatR}{{\boldsymbol{\mathsf{R}}}}
\safemath{\rmatS}{{\boldsymbol{\mathsf{S}}}}
\safemath{\rmatT}{{\boldsymbol{\mathsf{T}}}}
\safemath{\rmatU}{{\boldsymbol{\mathsf{U}}}}
\safemath{\rmatV}{{\boldsymbol{\mathsf{V}}}}
\safemath{\rmatW}{{\boldsymbol{\mathsf{W}}}}
\safemath{\rmatX}{{\boldsymbol{\mathsf{X}}}}
\safemath{\rmatY}{{\boldsymbol{\mathsf{Y}}}}
\safemath{\rmatZ}{{\boldsymbol{\mathsf{Z}}}}
\safemath{\rmatDelta}{\bmuDelta}
\safemath{\rmatGamma}{\bmuGamma}
\safemath{\rmatLambda}{\bmuLambda}
\safemath{\rmatOmega}{\bmuOmega}
\safemath{\rmatPhi}{\bmuPhi}
\safemath{\rmatPi}{\bmuPi}
\safemath{\rmatPsi}{\bmuPsi}
\safemath{\rmatSigma}{\bmuSigma}
\safemath{\rmatTheta}{\bmuTheta}
\safemath{\rmatUpsilon}{\bmuUpsilon}
\safemath{\rmatXi}{\bmuXi}

\safemath{\rmatAc}{\rnda}
\safemath{\rmatBc}{\rndb}
\safemath{\rmatCc}{\rndc}
\safemath{\rmatDc}{\rndd}
\safemath{\rmatEc}{\rnde}
\safemath{\rmatFc}{\rndf}
\safemath{\rmatGc}{\rndg}
\safemath{\rmatHc}{\rndh}
\safemath{\rmatIc}{\rndi}
\safemath{\rmatJc}{\rndj}
\safemath{\rmatKc}{\rndk}
\safemath{\rmatLc}{\rndl}
\safemath{\rmatMc}{\rndm}
\safemath{\rmatNc}{\rndn}
\safemath{\rmatOc}{\rndo}
\safemath{\rmatPc}{\rndp}
\safemath{\rmatQc}{\rndq}
\safemath{\rmatRc}{\rndr}
\safemath{\rmatSc}{\rnds}
\safemath{\rmatTc}{\rndt}
\safemath{\rmatUc}{\rndu}
\safemath{\rmatVc}{\rndv}
\safemath{\rmatWc}{\rndw}
\safemath{\rmatXc}{\rndx}
\safemath{\rmatYc}{\rndy}
\safemath{\rmatZc}{\rndz}

\newenvironment{textbmatrix}{	\setlength{\arraycolsep}{2.5pt}%
								\big[\begin{matrix}}{\end{matrix}\big]%
								\raisebox{0.08ex}{\vphantom{M}}}

 \def\btm{\begin{textbmatrix}}
 \def\etm{\end{textbmatrix}}

\DeclareMathOperator{\adj}{adj}				

\safemath{\fun}{\scf}						

\safemath{\vrbl}{x}						
\safemath{\altvrbl}{y}						
\safemath{\aaltvrbl}{z}						

\safemath{\vvrbl}{\vecx}						
\safemath{\altvvrbl}{\vecy}						
\safemath{\aaltvvrbl}{\vecz}						

\safemath{\altfun}{\scg}
\safemath{\aaltfun}{\sch}
\safemath{\bel}{\sce}					
\safemath{\altbel}{\sce}					
\safemath{\frel}{g}					
\safemath{\altfrel}{g}					
\safemath{\dfrel}{\tilde{g}}					
\safemath{\altdfrel}{\tilde{g}}					

\safemath{\mat}{\matA}						
\safemath{\matc}{\matAc}						
\safemath{\altmat}{\matB}						
\safemath{\altmatc}{\matBc}						

\safemath{\vectr}{\vecu}						
\safemath{\vectrc}{\vecuc}						
\safemath{\altvectr}{\vecv}						
\safemath{\aaltvectr}{\vect}						
\safemath{\altvectrc}{\vecvc}						
\safemath{\genvar}{u}						
\safemath{\altgenvar}{v}						

\safemath{\rvectr}{\rvecu}						
\safemath{\rvectrc}{\rvecuc}						
\safemath{\raltvectr}{\rvecv}						
\safemath{\raaltvectr}{\rvect}						
\safemath{\raltvectrc}{\rvecvc}						
\safemath{\rgenvar}{\rndu}						
\safemath{\raltgenvar}{\rndv}						

\newcommand{\nullspace}{\setN}	 			

\newcommand{\conj}[1]{\ensuremath{#1^{*}}} 	
\newcommand{\tp}[1]{\ensuremath{#1^{\mathsf{T}}}} 		

\newcommand{\inv}[1]{\ensuremath{#1^{-1}}} 	

\safemath{\dirac}{\delta}					
\safemath{\diracp}{\dirac(\time)}			
\safemath{\krond}{\dirac}					
\safemath{\indfun}{I}						
\safemath{\stepfun}{u}						

\safemath{\upto}{\uparrow}
\safemath{\downto}{\downarrow}
\safemath{\iu}{\mathrm{i}}							
\safemath{\maj}{\succ}
\newcommand{\dftmat}[1]{\matF_{#1}}			
\safemath{\mdft}{\dftmat{}}					
\safemath{\runity}{\beta}					
\safemath{\eval}{\lambda}					

\safemath{\veval}{\veclambda}				
\safemath{\littleo}{\sco}					

\let\im\undefined
\safemath{\re}{\Re}				
\safemath{\im}{\Im}				

\safemath{\euclidspace}{\complexset}			
\safemath{\confunspace}{\setC}				
\newcommand{\banachseqspace}[1]{l^{#1}}		
\safemath{\hilseqspace}{\banachseqspace{2}}	
\newcommand{\banachfunspace}[1]{\setL^{#1}}	
\safemath{\hilfunspace}{\banachfunspace{2}}	
\safemath{\hilfunspacep}{\hilfunspace(\complexset)}	
\safemath{\schwarzspace}{\setS}				

\newcommand{\hadj}[1]{#1^{\star}}			

\safemath{\SNR}{\rho} 				
\safemath{\SINR}{\text{\sc sinr}} 				
\safemath{\No}{N_0}							
\safemath{\Es}{E_s}							
\safemath{\Eb}{E_b}							
\safemath{\EbNo}{\frac{\Eb}{\No}}
\safemath{\EsNo}{\frac{\Es}{\No}}
\safemath{\NoVar}{\variance}                 

\let\time\undefined
\safemath{\time}{\sct}						
\safemath{\dtime}{\sck}						
\safemath{\delay}{\sctau}					
\safemath{\ddelay}{\scl}						
\safemath{\doppler}{\scnu}					
\safemath{\ddoppler}{\scm}					
\safemath{\freq}{\scf}						
\safemath{\dfreq}{\scn}						
\safemath{\Dtime}{\Delta\time}
\safemath{\Dfreq}{\Delta\freq}
\safemath{\Ddtime}{\dtime}
\safemath{\Ddfreq}{\dfreq}
\safemath{\bandwidth}{\scB}
\safemath{\maxdoppler}{\doppler_{0}}			
\safemath{\maxdelay}{\delay_{0}}				
\safemath{\spread}{\Delta_{\CHop}}			
\DeclareMathOperator{\CHop}{\ensuremath{\opH}} 
\safemath{\kernel}{\rndk_{\CHop}}			
\safemath{\kernelp}{\kernel(\time,\time')}	
\safemath{\tvir}{\rndh_{\CHop}}				
\safemath{\tvirp}{\tvir(\time,\delay)}		
\safemath{\tvirc}{\conj{\rndh}_{\CHop}}		
\safemath{\tvtf}{\rndl_{\CHop}}				
\safemath{\tvtfp}{\tvtf(\time,\freq)}			
\safemath{\tvtfc}{\conj{\rndl}_{\CHop}}		
\safemath{\spf}{\rnds_{\CHop}}				
\safemath{\spfp}{\spf(\doppler,\delay)}		
\safemath{\spfc}{\conj{\rnds}_{\CHop}}		
\safemath{\bff}{\rndb_{\CHop}}				
\safemath{\bffp}{\bff(\doppler,\freq)}		

\safemath{\irc}{\scr_{\rndh}}				
\safemath{\tfc}{\scr_{\rndl}}				
\safemath{\spc}{\scr_{\rnds}}				
\safemath{\bfc}{\scr_{\rndb}}				

\safemath{\scaf}{\scc_{\rnds}}				
\safemath{\scafp}{\scaf(\doppler,\delay)}		
\safemath{\ccf}{\scc_{\rndl}}				
\safemath{\ccfp}{\ccf(\Dtime,\Dfreq)}			
\safemath{\cic}{\scc_{\rndh}}				
\safemath{\cicp}{\cic(\Dtime,\delay)}			

\safemath{\mi}{I}							
\safemath{\capacity}{C}					

\DeclareMathOperator{\Prob}{\opP}		

\safemath{\normal}{\mathcal{N}}			
\safemath{\jpg}{\mathcal{CN}}			
\safemath{\uniform}{\mathcal{U}}				
\safemath{\mchain}{\leftrightarrow}		

\safemath{\dB}{\,\mathrm{dB}}
\safemath{\dBm}{\,\mathrm{dBm}}
\safemath{\Hz}{\,\mathrm{Hz}}
\safemath{\kHz}{\,\mathrm{kHz}}
\safemath{\MHz}{\,\mathrm{MHz}}
\safemath{\GHz}{\,\mathrm{GHz}}
\safemath{\s}{\,\mathrm{s}}
\safemath{\ms}{\,\mathrm{ms}}
\safemath{\mus}{\,\mathrm{\text{\textmu}s}}
\safemath{\ns}{\,\mathrm{ns}}
\safemath{\ps}{\,\mathrm{ps}}
\safemath{\meter}{\,\mathrm{m}}
\safemath{\mm}{\,\mathrm{mm}}
\safemath{\cm}{\,\mathrm{cm}}
\safemath{\m}{\,\mathrm{m}}
\safemath{\W}{\,\mathrm{W}}
\safemath{\mW}{\, \mathrm{mW}}
\safemath{\J}{\,\mathrm{J}}
\safemath{\K}{\,\mathrm{K}}
\safemath{\bit}{\,\mathrm{bit}}
\safemath{\nat}{\,\mathrm{nat}}

\safemath{\define}{\triangleq}					

\safemath{\equivalent}{\sim}
\safemath{\distas}{\sim}					
\safemath{\sdiff}{\Delta}				
\safemath{\setdiff}{\setminus}				

\safemath{\reals}{\mathbb R}
\safemath{\positivereals}{\reals^{+}}
\safemath{\integers}{\mathbb Z}
\safemath{\posint}{\integers^{+}}
\safemath{\naturals}{\mathbb N}
\safemath{\posnaturals}{\naturals^{+}}
\safemath{\complexset}{\mathbb C}
\safemath{\rationals}{\mathbb Q}

\safemath{\iSet}{\setI}
\safemath{\rel}{\bowtie}					
\safemath{\eqrel}{\sim}					
\safemath{\rlord}{\leq}					
\safemath{\slord}{<}						
\safemath{\rpord}{\preceq}				
\safemath{\rrpord}{\succeq}				
\safemath{\spord}{\prec}					
\safemath{\sig}{\sigma}					
\safemath{\metric}{d}					

\safemath{\setfun}{\Phi}					
\safemath{\measure}{\mu}					
\safemath{\altmeasure}{\lambda}					
\newcommand{\outerm}[1]{#1^{\star}}		
\newcommand{\innerm}[1]{#1_{\star}}		
\safemath{\omeasure}{\outerm{\measure}}		
\safemath{\imeasure}{\innerm{\measure}}		
\safemath{\aecol}{\colS^{\star}_{\measure}} 
\safemath{\emeasure}{\bar{\measure}_{0}}	
\safemath{\rmeasure}{\tilde{\measure}}	
\safemath{\bmeasure}{\measure_{0}}		
\safemath{\glength}{\measure_{\altfun}}	
\safemath{\lebmea}{\lambda}				
\safemath{\blebmea}{\lebmea_{0}}			
\safemath{\sfun}{s}						
\safemath{\absintspace}{\colL^{1}}		
\safemath{\sqintspace}{\colL^{2}}		
\safemath{\abssumspace}{l^{1}}		
\safemath{\sqsumspace}{l^{2}}		

\safemath{\sfield}{\setF}				
\safemath{\vectors}{\setV}				
\safemath{\vecspace}{(\vectors,\sfield)}	
\safemath{\linspace}{\setV}				
\safemath{\clinspace}{(\linspace,\sfield)} 
\safemath{\nspace}{\setU}				
\safemath{\metspace}{\setM}				
\safemath{\bspace}{\setB}				
\safemath{\ipspace}{\setG}				
\safemath{\hilspace}{\setH}				
\safemath{\blospace}{\setG}				
\safemath{\lop}{\opT}					
\safemath{\altlop}{\opS}					
\safemath{\nullsp}{\nullspace(\lop)}		
\safemath{\lfun}{l}						
\safemath{\altlfun}{g}					
\newcommand{\dual}[1]{#1^{'}}			
\safemath{\dsum}{\oplus}					
\safemath{\funspace}{\colL}				
\renewcommand{\adj}[1]{#1^{\times}}		
\safemath{\adjlop}{\adj{\lop}}			
\safemath{\hadjlop}{\hadj{\lop}}			
\safemath{\tow}{\xrightarrow{w}}			
\safemath{\tows}{\xrightarrow{w^{*}}}		
\safemath{\cparam}{\lambda}
\safemath{\lopl}{\lop_{\cparam}}		
\safemath{\iop}{\opI}					
\safemath{\resolop}{\opR}				
\safemath{\resolvent}{\resolop_{\cparam}(\lop)}	
\safemath{\reset}{\setQ}
\safemath{\spectrum}{\setS}
\safemath{\resolset}{\reset(\lop)}		
\safemath{\lopspec}{\spectrum(\lop)}		
\safemath{\pspec}{\spectrum_{p}(\lop)}	
\safemath{\cspec}{\spectrum_{c}(\lop)}	
\safemath{\rspec}{\spectrum_{r}(\lop)}	
\safemath{\ev}{\cparam}					
\newcommand{\specrad}[1]{r_{#1}}			
\safemath{\lopsrad}{\specrad{\lop}}		
\safemath{\pop}{\opP}					

\safemath{\specfam}{\colE}				
\safemath{\specop}{\opE_{\cparam}}		
\safemath{\altspecop}{\opE_{\mu}}		
\safemath{\vmulti}{\vecone}				
\safemath{\unitaryop}{\opU}				
\safemath{\sval}{\sigma}					
\safemath{\corrcoef}{\rho}				
\safemath{\sangle}{\theta}				

\let\time\undefined
\safemath{\iset}{\setI}				
\safemath{\shift}{\nu}
\safemath{\scale}{\alpha}
\safemath{\time}{t}
\safemath{\specfreq}{\theta}	
\newcommand{\transopgen}[1]{\opT_{#1}} 
\safemath{\transop}{\transopgen{\delay}}
\newcommand{\modopgen}[1]{\opM_{#1}}	
\safemath{\modop}{\modopgen{\shift}}
\newcommand{\dilopgen}[1]{\opD_{#1}}	
\safemath{\dilop}{\dilopgen{\scale}}
\safemath{\fram}{\setF}				
\safemath{\dfram}{\dual{\fram}}		
\safemath{\ufb}{B}					
\safemath{\lfb}{A}					
\safemath{\sop}{\hadj{\aop}}				
\safemath{\aop}{\opT}			
\safemath{\fop}{\opS}				
\safemath{\daop}{\tilde\opT}			
\safemath{\dsop}{\hadj{\tilde\opT}}				

\safemath{\ifop}{\inv{\fop}}			
\safemath{\rifop}{\fop^{-1/2}}			
\safemath{\transeq}{\setT}			
\safemath{\nfun}{\Phi}				
\safemath{\funvec}{\vecf}			
\safemath{\altfunvec}{\vecg}

\safemath{\samplespace}{\Omega}
\safemath{\probspace}{(\samplespace,\sfield,\Prob)}	
\safemath{\ccoef}{\rho}			

\safemath{\infstate}{\vecpi}				
\safemath{\typset}{\setA_{\epsilon}^{(N)}}	
\safemath{\expequal}{\doteq}				
\safemath{\code}{C}						
\safemath{\dstringset}{\setD^{\star}}		
\safemath{\cwlength}{l}					
\safemath{\elength}{L}					
\safemath{\extension}{C^{\star}}			
\safemath{\approaches}{\rightarrow}		
\safemath{\evnt}{\setA}					
\safemath{\altevnt}{\setB}					
\safemath{\rv}{\rndx}					
\safemath{\altrv}{\rndy}					
\safemath{\complexrv}{\rndu}					
\safemath{\altcrv}{\rndv}				
\safemath{\rvec}{\rvecx}					
\safemath{\altrvec}{\rvecy}				
\safemath{\crvec}{\rvecu}				
\safemath{\altcrvec}{\rvecv}				
\safemath{\variance}{\sigma^{2}}			
\safemath{\map}{T}						
\safemath{\jacobian}{\matJ}					
\safemath{\wvec}{\rvecw}					
\safemath{\wrv}{\rndw}					
\safemath{\orthmat}{\matQ}				
\safemath{\evmat}{\matLambda}			
\safemath{\identity}{\matidentity}		
\safemath{\innovec}{\vecv}				
\safemath{\convas}{\xrightarrow{\text{a.s.}}}	
\safemath{\convr}{\xrightarrow{\text{r}}}	
\safemath{\convp}{\xrightarrow{\text{P}}}	
\safemath{\convd}{\xrightarrow{\text{D}}}	
\safemath{\ltis}{\opL}				
\safemath{\ir}{h}					
\safemath{\tf}{\MakeUppercase{\ir}}	

\newcommand*{\fancyrefparlabelprefix}{par}		
\newcommand*{\fancyrefremlabelprefix}{rem}		
\newcommand*{\fancyrefchalabelprefix}{cha}		
\newcommand*{\fancyrefapplabelprefix}{app}		
\newcommand*{\fancyrefthmlabelprefix}{thm}		
\newcommand*{\fancyreflemlabelprefix}{lem}		
\newcommand*{\fancyrefcorlabelprefix}{cor}		
\newcommand*{\fancyrefdeflabelprefix}{def}		
\newcommand*{\fancyrefproplabelprefix}{prop}		
\frefformat{vario}{\fancyrefparlabelprefix}{Part~#1}
\frefformat{vario}{\fancyrefremlabelprefix}{Remark~#1}
\frefformat{vario}{\fancyrefchalabelprefix}{Chapter~#1}
\frefformat{vario}{\fancyrefseclabelprefix}{Section~#1}

\frefformat{vario}{\fancyrefthmlabelprefix}{Theorem~#1}
\frefformat{vario}{\fancyreflemlabelprefix}{Lemma~#1}
\frefformat{vario}{\fancyrefcorlabelprefix}{Corollary~#1}
\frefformat{vario}{\fancyrefdeflabelprefix}{Definition~#1}
\frefformat{vario}{\fancyreffiglabelprefix}{Figure~#1}
\frefformat{vario}{\fancyrefapplabelprefix}{Appendix~#1}
\frefformat{vario}{\fancyrefeqlabelprefix}{(#1)}
\frefformat{vario}{\fancyrefproplabelprefix}{Property~#1}

\theoremstyle{plain}
\newtheorem{thm}{Theorem}
\newtheorem{lem}{Lemma}
\newtheorem{prp}{Proposition}
\newtheorem{cor}{Corollary}
\theoremstyle{definition}

\theoremstyle{remark}

\newtheorem{exa}{Example}
\newtheorem{dfn}{Definition}

